\documentclass[aps,prd,onecolumn,showpacs,showkeys,amsmath,amssymb]{revtex4}
\usepackage{epsfig} %%% ÎÄÖÐ²åÈë eps ÍŒÐÎÊ±Ò²¿Éµ÷ÓÃŽËºê°ü£¬²åÍŒ·œ·š 2
\usepackage{amsmath}  %%% ÅÅÓ¡±ÈœÏžŽÔÓµÄÊýÑ§¹«ÊœºÍ·ûºÅÊ±¿Éµ÷ÓÃŽËºê°ü
\usepackage{makecell}
\usepackage{graphicx}
\usepackage{array}
\usepackage{subfigure}
\usepackage{bm}
\usepackage{longtable}
\usepackage{booktabs}
\usepackage{appendix}
\usepackage{amsfonts}
\usepackage{amsmath}
\usepackage{dcolumn}
\usepackage{siunitx}
\usepackage{bm}
\usepackage{booktabs}
\usepackage[utf8]{inputenc}

\usepackage{float}
\usepackage{longtable,lscape}
\usepackage{txfonts}
\usepackage{overpic}
\usepackage{amssymb}
\usepackage{indentfirst}
\usepackage{epsfig}
\usepackage{feynmf}   %{feynmp}
\usepackage{epstopdf}   %{feynmp}
\usepackage{slashed}  %for Feynman symbols
\usepackage{multirow}
\usepackage[colorlinks, citecolor=blue,anchorcolor=red,menucolor=red, linkcolor=red,filecolor=red,runcolor=red,urlcolor=blue,frenchlinks=red]{hyperref}

\makeatletter
\newcommand{\figcaption}{\def\@captype{figure}\caption}
\newcommand{\tabcaption}{\def\@captype{table}\caption}

\newcommand{\Rmnum}[1]{\expandafter\@slowromancap\romannumeral #1@}
\def\hlinewd#1{%
  \noalign{\ifnum0=`}\fi\hrule \@height #1 \futurelet
   \reserved@a\@xhline}
\makeatother

\begin{document}
\title{\vspace{-30mm}Mass spectra of strange double charm pentaquarks with strangeness $S=-1$\vspace{9mm}}

\author{Zi-Yan Yang$^{1,2}$}
\author{Qian Wang$^{1,2,4}$}
\email{qianwang@m.scnu.edu.cn}
\author{Wei Chen$^{3,4}$}
\email{chenwei29@mail.sysu.edu.cn}

\affiliation{$^1$Key Laboratory of Atomic and Subatomic Structure and Quantum Control (MOE), Guangdong Basic Research Center of Excellence for Structure and Fundamental Interactions of Matter, Institute of Quantum Matter, South China Normal University, Guangzhou 510006, China}
\affiliation{$^2$Guangdong-Hong Kong Joint Laboratory of Quantum Matter, Guangdong Provincial Key Laboratory of Nuclear Science, Southern Nuclear Science Computing Center, South China Normal University, Guangzhou 510006, China}
\affiliation{$^3$School of Physics, Sun Yat-sen University, Guangzhou 510275, China}
\affiliation{$^4$Southern Center for Nuclear-Science Theory (SCNT), Institute of Modern Physics, 
Chinese Academy of Sciences, Huizhou 516000, Guangdong Province, China}

\begin{abstract}
The observation of the $T_{c\bar{s}}(2900)$ indicates the potential existence of strange double charm pentaquarks based on the heavy antidiquark symmetry. We systematically study 
the mass spectra of strange double charm pentaquarks with strangeness $S=-1$ in both molecular and compact structures for quantum numbers $J^{P}=1/2^{-}$, $3/2^{-}$, $5/2^{-}$. By constructing the interpolating currents, the mass spectra can be extracted from the two-point correlation functions in the framework of QCD sum rule method. In the molecular picture, we 
 find that the $\Xi_c^+D^{\ast +}$, $\Xi_c^{'+}D^{\ast +}$, $\Xi_{c}^{\ast +}D^{\ast +}$, $\Xi_{cc}^{\ast ++}\bar{K}^{\ast 0}$ and $\Omega_{cc}^{\ast +}\rho^{+}$ may form molecular strange double charm pentaquarks. In both pictures, the masses of the $J^P=1/2^-, 3/2^-$ pentaquarks locate within the $4.2-4.6~\mathrm{GeV}$ and $4.2-4.5~\mathrm{GeV}$ regions, respectively. As all of them are above the thresholds of their strong decay channels, they behave as a broad state, making them challenging to be detected in experiment. On the contrary, 
 the mass of the $J^P=5/2^-$ strange double charm pentaquark is located at $4.3~\mathrm{GeV}$ and below its strong decay channel. This makes it as a narrow state and easy to be identified in experiment. The best observed channel is its semi-leptonic decay to double charm baryon. As the result, we strongly suggest experiments to search for $J^P=5/2^-$ strange double charm pentaquarks as a first try.
\end{abstract}
\maketitle

%\tableofcontents
\pagenumbering{arabic}
\section{Introduction}\label{Sec:Intro}
\par The study of multiquarks goes back to 
half a century after the quark model proposed in 1964 ~\cite{Gell-Mann:1964ewy,1964-Zweig-p-} 
and becomes a hot topic since the observation of the $X(3872)$ in 2003. Due to the sufficient statistic in experiment, tens of charmonium-like and bottomonium-like states are observed by various experimental collaborations~\cite{ParticleDataGroup:2022pth}.
These states could be beyond the conventional quark model and be viewed as exotic candidates.
They also provide a novel platform to
shed light on the hadronization mechanism.
Although numerous theoretical efforts have been put forward to understand their nature~\cite{Nielsen:2009uh,Chen:2016qju,Richard:2016eis,Esposito:2016noz,Ali:2017jda,Guo:2017jvc,Albuquerque:2018jkn,Liu:2019zoy,Brambilla:2019esw,Richard:2019cmi,Faustov:2021hjs,Chen:2022asf,Meng:2022ozq}, it is still unclear about the hadronization mechanism.

\par Because of the observation of these exotic 
candidates, in a more general concept, all the 
bosons are defined as mesons and all the fermions are defined as baryons. The latter one is more complicated than the former one due to one additional (anti)quark. The first well-established exotic baryon signal is reported by the LHCb collaboration in the $J/\psi p$ invariant mass distribution of the $\Lambda_b^0\rightarrow J/\psi K^-p$ process~\cite{LHCb:2015yax}. 
The two structures are named as $P_c(4380)$ and $P_c(4450)$. Four years later, with an order-of-magnitude larger statistic data, the LHCb collaboration further reported their hyperfine 
structures~\cite{LHCb:2019kea}. The $P_c(4380)$
 is split into two structures $P_c(4440)$ and $P_c(4457)$ and a new narrow peak $P_c(4312)$
 emerges. In 2020, the LHCb collaboration observed the strange partner, i.e.the $P_{cs}(4459)$ state, of the $P_c$ states
 in the $J/\psi\Lambda$ invariant mass distribution of the $\Xi_b^-\rightarrow J/\psi\Lambda K^-$ process~\cite{LHCb:2020jpq}. 
 Very recently, another very narrow resonance $P_{cs}(4338)$ is reported in the $J/\psi\Lambda$ invariant mass of the $B^-\rightarrow J/\psi\Lambda \bar{p}$ process, with the preferred $J^P=1/2^-$ at $90\%$ confidence level~\cite{LHCb:2022ogu}.
 Due to their observed channel, the quark contents of $P_c$ and $P_{cs}$ are $uudc\bar{c}$ and $udsc\bar{c}$  respectively, 
 indicating they are hidden charm pentaquarks.
 
 The recent discovery of double charm tetraquark $T_{cc}^+(3875)$~\cite{LHCb:2021vvq,LHCb:2021auc} raises a question whether double charm pentaquarks exist or not. Some theoretical attempts have made for considering the mass spectrum from the hadronic molecular picture~\cite{Dong:2021bvy,Chen:2017vai,Guo:2017vcf,Zhu:2019iwm,Wang:2023eng,Duan:2024uuf} and the compact pentaquark picture~\cite{Chen:2021kad,Xing:2021yid,Zhou:2018bkn,Wang:2018lhz,Park:2018oib}. Some theoretical attempts have made for considering the electromagnetic properties~\cite{Ozdem:2022vip, Ozdem:2024yel}. For the double charm pentaquarks,
 the observed $\Xi_{cc}$~\cite{LHCb:2017iph,LHCb:2018pcs} provide an important input from the experimental side. 
 In Ref.\cite{Dong:2021bvy}, the authors work on a Bethe-Salpeter equation with the interaction respecting heavy quark spin symmetry and predict a $D^{(*)}\Xi_{c}^{('*)}$ bound state.  Ref.\cite{Zhou:2018bkn} works in color-magnetic interaction and predicts several compact $QQqq\bar{q}$ states,
 which could be searched in the $\Omega_{cc}\pi$, $\Xi_{cc}K$ and $\Xi_cD$ channels. Ref.\cite{Xing:2021yid}
 performs a study within the double heavy triquark-diquark framework respecting SU(3) flavor symmetry. 
 The study finds several stable double charm pentaquarks, for instance a $J^P=1/2^-$ $cc\bar{s}ud$ double charm pentaqurk,
 against their strong decay channels. Ref.~\cite{Duan:2024uuf} considers the potential double charm pentaquarks $P_{cc}$ with quark content $ccud\bar{d}$ in QCD sum rule method. In this work, we further consider the double charm pentaquarks $P_{cc}$ with quark content $ccus\bar{d}$,
 based on the heavy antiquark-diquark symmetry (HADS) proposed in Ref.~\cite{Savage:1990di}. The HADS states 
 that a color triplet double heavy quarks behaves like a heavy antiquark in color space. In this case, the observed
 $T_{c\bar{s}0}^a(2900)^0$~\cite{LHCb:2022lzp} with quark content $cd\bar{u}\bar{s}$ indicates the potential existence of strange double charm pentaquark $ccus\bar{d}$. 
 From the HADS point of view, the mass of double heavy pentaquark satisfies the relation
\begin{equation}\label{Eq:HADSPcc}
m(QQqq\bar{q})-m(QQ\bar{q}\bar{q})=m(qq\bar{q}\bar{Q})-m(\bar{Q}\bar{q}\bar{q}),
\end{equation}
by replacing $\bar{Q}\rightarrow QQ$, as two heavy quarks in color antitriplet behave like a steady color source from a heavy antiquark. The essence of Eq.~\eqref{Eq:HADSPcc} can be traced back to Refs.~\cite{Eichten:1987xu,Lepage:1987gg}. From this point of view, the recently observed $T_{cc}^+(3875)$, as a isospin singlet state, is related to the $\bar{\Lambda}_c$ by the HADS. The $T_{c\bar{s}0}^{0}(2900)$ indicates the existence of strange double charm pentaquark. 

\par Based on the above arguments, we shall systematically study double charm pentaquarks with quark content $ccus\bar{d}$. To obtain a solid conclusion, we start from both the hadronic molecular currents, i.e. $\Xi_c^+D^+$, $\Xi_{cc}^{++}\bar{K}^0$, $\Omega_{cc}^{+}\pi^+$, and the compact pentaquark currents in QCD sum rule approach\cite{Reinders:1984sr,Shifman:1978bx}. 
 The paper is organized as follows. In Sec.\ref{Sec:Current}, we construct the local pentaquark interpolating currents for double heavy pentaquark states. Using these currents, we perform the parity-projected QCD sum rules analysis in Sec.\ref{Sec:QCDSR}.  The numerical analysis follows in Sec.\ref{Numerical}. The results and discussions are presented in Sec.\ref{Resultanddis}.

\section{Interpolating Current for double Heavy Pentaquark}\label{Sec:Current}
\par In this section we systematically construct the local pentaquark interpolating currents with spin-parity $J^P=1/2^-,3/2^-,5/2^-$, since these quantum numbers can be achieved with the $S$-wave ground heavy (or double heavy) baryon and ground charm (or light) meson with quark content $QQus\bar{d}$. Five flavor 
configurations $[\bar{d}_ds_d][\epsilon^{abc}Q_aQ_bu_c]$ and $[\bar{d}_dQ_d][\epsilon^{abc}Q_au_bs_c]$, $[\bar{d}_du_d][\epsilon^{abc}Q_aQ_bs_c]$, $\epsilon^{aij}\epsilon^{bkl}\epsilon^{abc}[Q_iu_j][Q_ks_l]\bar{d}_c$ and $\epsilon^{aij}\epsilon^{bkl}\epsilon^{abc}[Q_iQ_j][u_ks_l]\bar{d}_c$ are considered.  Here $a\cdots d$, $i\cdots l$ are color indices, $u,d,s$ represents the up, down and strange quark, $Q$ represents the heavy quark, i.e. charm or bottom quark.  The former three flavor configurations have the same color configuration $\mathbf{1}_c\otimes\mathbf{1}_c$, which can be related by the famous Fierz transformation:
\begin{equation}
\delta^{de}\epsilon^{abc}=\delta^{da}\epsilon^{ebc}+\delta^{db}\epsilon^{aec}+\delta^{dc}\epsilon^{abe}.
\end{equation}
The latter two flavor configurations have the same color configuration $\bar{\mathbf{3}}_c\otimes\bar{\mathbf{3}}_c\otimes\bar{\mathbf{3}}_c$.

\par In order to find the correspondence for open charm tetraquark $us\bar{d}\bar{c}$ and double charm pentaquark $ccus\bar{d}$ due to HADS, we analyse their spin structure. Insuring the charm diquark has the same color structure, i.e. antisymmetric in color space, as anti-charm quark, the spin structure of diquark should be symmetric due to Pauli principle, which requires the spin of charm diquark should be $S_{[cc]}=1$. As the strange charm tetraquark $T_{c\bar{s}0}^{0}(2900)$ with quark content $[u\bar{c}][s\bar{d}]$ is a spin singlet state~\cite{LHCb:2022lzp},
 the spin structure of diquark and antidiquark should be $[u\bar{c}]_{\mathbf{0}}[s\bar{d}]_{\mathbf{0}}$ or $[u\bar{c}]_{\mathbf{1}}[s\bar{d}]_{\mathbf{1}}$ to form a spin-0 tetraquark state. Thus for open charm tetraquark with $J^P=0^+$, the spin structure of corresponding HADS pentaquark partner should be 
\begin{equation*}
\mathbf{1}_{[cc]}\otimes\mathbf{\frac{1}{2}}_{[u]}\otimes\mathbf{0}_{[s\bar{d}]}=\mathbf{\frac{1}{2}}_{[ccus\bar{d}]}\oplus\mathbf{\frac{3}{2}}_{[ccus\bar{d}]}
\end{equation*}
doublet with spin-0 $[s\bar{d}]$ component or
\begin{equation*}
\mathbf{1}_{[cc]}\otimes\mathbf{\frac{1}{2}}_{[u]}\otimes\mathbf{1}_{[s\bar{d}]}=\mathbf{\frac{1}{2}}_{[ccus\bar{d}]}\oplus\mathbf{\frac{3}{2}}_{[ccus\bar{d}]}\oplus\mathbf{\frac{5}{2}}_{[ccus\bar{d}]}
\end{equation*}
triplet with spin-1 $[s\bar{d}]$ component, which indicates that the HADS partner for open charm tetraquark with $J^P=0^+$ should have spin $1/2,3/2$ or $5/2$. Analogous discussion can also be made for the molecular structure $[s\bar{c}][u\bar{d}]$ and the compact structure $[us][\bar{c}\bar{d}]$. 
 With the above considerations, in this work, we discuss various types of pentaquarks with spin $1/2,3/2$ and $5/2$.
 
\subsection{Currents for the heavy baryon-heavy meson molecular pentaquark}
\par The currents with color configuration $\epsilon_{abc}[u_as_bQ_c][\bar{d}_dQ_d]$ could well couple to $\Xi_c^{+}D^{+(*)}$ molecular states. In previous QCD sum rules analysis~\cite{Zhang:2009iya}, one finds that the currents
\begin{eqnarray}
&\frac{1}{\sqrt{2}}\epsilon_{abc}\left[\left(u_a^TC\gamma_5s_b-s_a^TC\gamma_5u_b\right)c_c\right],\\
&\frac{1}{\sqrt{2}}\epsilon_{abc}\left[\left(u_a^TC\gamma_\mu\gamma_5s_b-s_a^TC\gamma_\mu\gamma_5u_b\right)\gamma_\mu c_c\right],\\
&\sqrt{\frac{2}{3}}\epsilon_{abc}\left[(s^T_aC\gamma_\mu u_b)\gamma_5c_c+(u^T_aC\gamma_\mu c_b)\gamma_5s_c+(c^T_aC\gamma_\mu s_b)\gamma_5u_c\right],
\end{eqnarray}
can well couple to $\Xi_c^+$, $\Xi_c^{'+}$ and $\Xi_{c}^{\ast +}$ states, respectively. Here $\Xi_c^+$, $\Xi_c^{\prime +}$, $\Xi_c^{\ast +}$ belong to the SU(4) flavor spin-$\frac{1}{2}$  20-plet, spin-$\frac{1}{2}$ 20-plet, spin-$\frac{3}{2}$ 20-plet, respectively. Thus we can construct the following currents to perform QCD sum rule analysis:
\allowdisplaybreaks{
\begin{eqnarray}\label{Eq:eta}
\nonumber \eta_1&=&\frac{1}{\sqrt{2}}\epsilon_{abc}\left[\left(u_a^TC\gamma_5s_b-s_a^TC\gamma_5u_b\right)Q_c\right]\left[\bar{d}_d\gamma_5Q_d\right],\\
\nonumber \eta_2&=&\frac{1}{\sqrt{2}}\epsilon_{abc}\left[\left(u_a^TC\gamma_\mu\gamma_5s_b-s_a^TC\gamma_\mu\gamma_5u_b\right)\gamma_\mu Q_c\right]\left[\bar{d}_d\gamma_5Q_d\right],\\
\nonumber \eta_{3}&=&\frac{1}{\sqrt{2}}\epsilon_{abc}\left[\left(u_a^TC\gamma_5s_b-s_a^TC\gamma_5u_b\right)\gamma_\mu Q_c\right]\left[\bar{d}_d\gamma_\mu Q_d\right],\\
\eta_{4\mu}&=&\frac{1}{\sqrt{2}}\epsilon_{abc}\left[\left(u_a^TC\gamma_\nu\gamma_5s_b-s_a^TC\gamma_\nu\gamma_5u_b\right)\gamma_\nu Q_c\right]\left[\bar{d}_d\gamma_\mu Q_d\right],\\
\nonumber \eta_{5\mu}&=&\sqrt{\frac{2}{3}}\epsilon_{abc}\left[(s^T_aC\gamma_\mu u_b)\gamma_5Q_c+(u^T_aC\gamma_\mu Q_b)\gamma_5s_c+(Q^T_aC\gamma_\mu s_b)\gamma_5u_c\right]\left[\bar{d}_d\gamma_5Q_d\right],\\
\nonumber \eta_{6}&=&\sqrt{\frac{2}{3}}\epsilon_{abc}\left[(s^T_aC\gamma_\mu u_b)\gamma_5Q_c+(u^T_aC\gamma_\mu Q_b)\gamma_5s_c+(Q^T_aC\gamma_\mu s_b)\gamma_5u_c\right]\left[\bar{d}_d\gamma_\mu Q_d\right],\\
\nonumber \eta_{7,\mu\nu}&=&\sqrt{\frac{2}{3}}\epsilon_{abc}\left[(s^T_aC\gamma_\mu u_b)\gamma_5Q_c+(u^T_aC\gamma_\mu Q_b)\gamma_5 s_c+(Q^T_aC\gamma_\mu s_b)\gamma_5 u_c\right]\left[\bar{d}_d\gamma_\nu Q_d\right]+(\mu\leftrightarrow\nu),
\end{eqnarray}}
where $u,d,s$ denote up, down and strange quarks, respectively. Here $Q$ denotes heavy quarks $c$ or $b$. $T$ denotes the transpose of quark field. $C$ denotes the charge conjugation operator. The indexes $a,b,c,d$ are the color indices of quark fields. 
One notices that not all of the above currents are related to the strange charm tetraquark by HADS. Further discussions can be found in Sec.~\ref{Resultanddis}. 

\par It should be noted that, the interpolating currents for baryon states could couple to both positive and negative parity states, thus currents in Eq.~\eqref{Eq:eta} could couple to $J^P=1/2^\pm,3/2^\pm$ or $5/2^\pm$ pentaquarks. For instance, the current $\eta_1$ would couple to both $\Xi_c^+D^+$ states with $J^P=1/2^-$, and $\Xi_{c}^+D^+$ states with $J^P=1/2^+$ in $P$-wave. We will further discuss such an issue in the next section.

\subsection{Currents for the double heavy baryon-light meson molecular pentaquark}
\par We introduce currents with color configuration $\epsilon_{abc}[Q_aQ_bu_c][\bar{d}_ds_d]$ and $\epsilon_{abc}[Q_aQ_bs_c][\bar{d}_du_d]$ coupling to $\Xi_{cc}^{++}\bar{K}^{0(\ast)}$ and $\Omega_{cc}^{+}\pi^{+}(\rho^{+})$ molecular states, respectively. In previous QCD sum rules analysis~\cite{Zhang:2009iya}, one suggests that the  currents
\begin{eqnarray}
&\epsilon_{abc}(c^T_aC\gamma_\mu c_b)\gamma_\mu\gamma_5u_c,\\
&\frac{1}{\sqrt{3}}\epsilon_{abc}\left[2\left(u^T_aC\gamma_\mu c_b\right)\gamma_5c_c+\left(c^T_aC\gamma_\mu c_b\right)\gamma_5u_c\right],
\end{eqnarray}
could well couple to $\Xi_{cc}^{++}$ and $\Xi_{cc}^{*++}$ states with $J^P=\frac{1}{2}^+,\frac{3}{2}^+$, respectively. Thus we can construct the following pentaquark currents to perform QCD sum rules analysis:
\allowdisplaybreaks{
\begin{eqnarray}\label{Eq:xi}
\nonumber \xi_1&=&\left[\epsilon_{abc}(Q^T_aC\gamma_\mu Q_b)\gamma_\mu\gamma_5u_c\right]\left[\bar{d}_d\gamma_5s_d\right],\\
\nonumber \xi_{2\mu}&=&\left[\epsilon_{abc}(Q^T_aC\gamma_\nu Q_b)\gamma_\nu\gamma_5u_c\right]\left[\bar{d}_d\gamma_\mu s_d\right],\\
\xi_{3\mu}&=&\frac{1}{\sqrt{3}}\epsilon_{abc}\left[2\left(u^T_aC\gamma_\mu Q_b\right)\gamma_5Q_c+\left(Q^T_aC\gamma_\mu Q_b\right)\gamma_5u_c\right]\left[\bar{d}_d\gamma_5s_d\right],\\
\nonumber \xi_{4}&=&\frac{1}{\sqrt{3}}\epsilon_{abc}\left[2\left(u^T_aC\gamma_\mu Q_b\right)\gamma_5Q_c+\left(Q^T_aC\gamma_\mu Q_b\right)\gamma_5u_c\right]\left[\bar{d}_d\gamma_\mu s_d\right],\\
\nonumber \xi_{5,\mu\nu}&=&\frac{1}{\sqrt{3}}\epsilon_{abc}\left[2\left(u^T_aC\gamma_\mu Q_b\right)\gamma_5Q_c+\left(Q^T_aC\gamma_\mu Q_b\right)\gamma_5u_c\right]\left[\bar{d}_d\gamma_\nu s_d\right]+(\mu\leftrightarrow\nu),
\end{eqnarray}}
where $\xi_i$ should be the HADS partner of open heavy tetraquark $[u\bar{c}]_{\mathbf{0}(\mathbf{1})}[s\bar{d}]_{\mathbf{0}(\mathbf{1})}$ due to its spin-1 $[cc]$ diquark component and spin-0(1) $[s\bar{d}]$ component. 
\par The interpolating currents for configuration $\epsilon_{abc}[Q_aQ_bs_c][\bar{d}_du_d]$ are the same as $\epsilon_{abc}[Q_aQ_bu_c][\bar{d}_ds_d]$ with substitution $u\leftrightarrow s$ and we denote them as $\psi_i$:
\begin{equation}\label{Eq:psi}
\psi_i=\xi_i\;(u\leftrightarrow s),
\end{equation}
where $\psi_i$ should be the HADS partner of open heavy tetraquark $[s\bar{c}]_{\mathbf{0}(\mathbf{1})}[u\bar{d}]_{\mathbf{0}(\mathbf{1})}$ due to its spin-1 $[cc]$ diquark component and spin-0(1) $[u\bar{d}]$ component. 

\subsection{Currents for compact pentaquark}
\par The double heavy pentaquarks with compact structure can be treated by an intuitive picture, in which the heavier component forms a nucleus and the lighter one is in an orbit around this nucleus. Such a picture for compact pentaquark has color configuration $[[cc]_{\bar{\mathbf{3}}}\bar{d}_{\bar{\mathbf{3}}}]_{\mathbf{3}}[us]_{\bar{\mathbf{3}}}$, where the heavy diquark forms a color triplet spin-1 state, as suggested previously. Since the $[cc]$ diquark with spin-0 violates the Fermi-Dirac statistic, we only construct the pentaquark with the $[cc]$ diquark with spin-1. To compare with this compact picture, we also consider a more complicated picture, i.e. one heavy quark and one light quark form one color antitriplet diquark and the two heavy-light diquarks combine the other light antiquark to form a color singlet pentaquark. 
 Such a picture has color configuration $[[cu]_{\bar{\mathbf{3}}}[cs]_{\bar{\mathbf{3}}}]_{\mathbf{3}}\bar{d}_{\bar{\mathbf{3}}}$. In this work, we suggest currents with color configuration $\epsilon^{aij}\epsilon^{bkl}\epsilon^{abc}[Q_iu_j][Q_ks_l]\bar{d}_c$ and $\epsilon^{aij}\epsilon^{bkl}\epsilon^{abc}[Q_iQ_j][u_ks_l]\bar{d}_c$ coupling to the two compact pentaquark states above, and we use the following interpolating currents to perform our QCD sum rule analysis:
\allowdisplaybreaks{
\begin{eqnarray}\label{Eq:J1}
%\nonumber J_{1,1}&=&\epsilon_{aij}\epsilon_{bkl}\epsilon_{abc}\left(Q_i^TC\gamma_5Q_j\right)\left(u_k^TC\gamma_5s_l\right)\gamma_5C\bar{d}_c^T,\\
\nonumber J_{1,2}&=&\epsilon_{aij}\epsilon_{bkl}\epsilon_{abc}\left(Q_i^TC\gamma_\mu Q_j\right)\left(u_k^TC\gamma_\mu s_l\right)\gamma_5C\bar{d}_c^T,\\
\nonumber J_{1,3}&=&\epsilon_{aij}\epsilon_{bkl}\epsilon_{abc}\left(Q_i^TC\gamma_\mu Q_j\right)\left(u_k^TC\gamma_5s_l\right)\gamma_\mu C\bar{d}_c^T,\\
%J_{1,4}&=&\epsilon_{aij}\epsilon_{bkl}\epsilon_{abc}\left(Q_i^TC\gamma_5Q_j\right)\left(u_k^TC\gamma_\mu s_l\right)\gamma_\mu C\bar{d}_c^T,\\
\nonumber J_{1,5\mu}&=&\epsilon_{aij}\epsilon_{bkl}\epsilon_{abc}\left(Q_i^TC\gamma_\mu Q_j\right)\left(u_k^TC\gamma_5 s_l\right)\gamma_5 C\bar{d}_c^T,\\
%\nonumber J_{1,6\mu}&=&\epsilon_{aij}\epsilon_{bkl}\epsilon_{abc}\left(Q_i^TC\gamma_5Q_j\right)\left(u_k^TC\gamma_\mu s_l\right)\gamma_5 C\bar{d}_c^T,\\
%\nonumber J_{1,7\mu}&=&\epsilon_{aij}\epsilon_{bkl}\epsilon_{abc}\left(Q_i^TC\gamma_5Q_j\right)\left(u_k^TC\gamma_5 s_l\right)\gamma_\mu C\bar{d}_c^T,\\
\nonumber
J_{1,8\mu}&=&\epsilon_{aij}\epsilon_{bkl}\epsilon_{abc}\left(Q_i^TC\gamma_\nu Q_j\right)\left(u_k^TC\gamma_\nu s_l\right)\gamma_\mu C\bar{d}_c^T,\\
\nonumber J_{1,9\mu\nu}&=&\epsilon_{aij}\epsilon_{bkl}\epsilon_{abc}\left(Q_i^TC\gamma_\mu Q_j\right)\left(u_k^TC\gamma_\nu s_l\right)\gamma_5 C\bar{d}_c^T+(\mu\leftrightarrow\nu),
\end{eqnarray}}
and
\allowdisplaybreaks{
\begin{eqnarray}\label{Eq:J2}
\nonumber J_{2,1}&=&\epsilon_{aij}\epsilon_{bkl}\epsilon_{abc}\left(Q_i^TC\gamma_5u_j\right)\left(Q_k^TC\gamma_5s_l\right)\gamma_5C\bar{d}_c^T,\\
\nonumber J_{2,2}&=&\epsilon_{aij}\epsilon_{bkl}\epsilon_{abc}\left(Q_i^TC\gamma_\mu u_j\right)\left(Q_k^TC\gamma_\mu s_l\right)\gamma_5C\bar{d}_c^T,\\
\nonumber J_{2,3}&=&\epsilon_{aij}\epsilon_{bkl}\epsilon_{abc}\left(Q_i^TC\gamma_\mu u_j\right)\left(Q_k^TC\gamma_5s_l\right)\gamma_\mu C\bar{d}_c^T,\\
J_{2,4}&=&\epsilon_{aij}\epsilon_{bkl}\epsilon_{abc}\left(Q_i^TC\gamma_5u_j\right)\left(Q_k^TC\gamma_\mu s_l\right)\gamma_\mu C\bar{d}_c^T,\\
\nonumber J_{2,5\mu}&=&\epsilon_{aij}\epsilon_{bkl}\epsilon_{abc}\left(Q_i^TC\gamma_\mu u_j\right)\left(Q_k^TC\gamma_5 s_l\right)\gamma_5 C\bar{d}_c^T,\\
\nonumber J_{2,6\mu}&=&\epsilon_{aij}\epsilon_{bkl}\epsilon_{abc}\left(Q_i^TC\gamma_5u_j\right)\left(Q_k^TC\gamma_\mu s_l\right)\gamma_5 C\bar{d}_c^T,\\
\nonumber J_{2,7\mu}&=&\epsilon_{aij}\epsilon_{bkl}\epsilon_{abc}\left(Q_i^TC\gamma_5u_j\right)\left(Q_k^TC\gamma_5 s_l\right)\gamma_\mu C\bar{d}_c^T,\\
\nonumber J_{2,8\mu\nu}&=&\epsilon_{aij}\epsilon_{bkl}\epsilon_{abc}\left(Q_i^TC\gamma_\mu u_j\right)\left(Q_k^TC\gamma_\nu s_l\right)\gamma_5 C\bar{d}_c^T+(\mu\leftrightarrow\nu),
\end{eqnarray}}
where currents $J_{1,i}$ are heavy diquark coupled and $J_{2,i}$ are heavy diquark decoupled. For convenience, we call these two types of currrents Type-I and Type-II. The currents $J_{1,2},J_{1,3},J_{1,5\mu},J_{1,8\mu},J_{1,9\mu\nu}$ should couple to the HADS partners of open heavy tetraquark $[us\bar{d}\bar{c}]$ due to their spin-1 $[cc]$ diquark components.

\section{QCD Sum Rules}\label{Sec:QCDSR}
\par In this section, we shall investigate the currents using the method of QCD sum rules. Symbols $J$, $J_\mu$ and $J_{\mu\nu}$ are assigned to denote the currents with spin $J=1/2,3/2,5/2$. The two-point correlation functions obtained by the currents can be written as \cite{Wang:2015epa}
\allowdisplaybreaks{
\begin{eqnarray}\label{correlation}
\nonumber\Pi(q^2)&=&i\int d^4x e^{iq\cdot x}\langle 0|T\left[J(x)\bar{J}(0)\right]|0\rangle\\
\nonumber&=&(\slashed{q}+M_X)\Pi^{1/2}(q^2),\\
\Pi_{\mu\nu}(q^2)&=&i\int d^4x e^{iq\cdot x}\langle 0|T\left[J_\mu(x)\bar{J}_\nu(0)\right]|0\rangle\\
\nonumber&=&\left(\frac{q_\mu q_\nu}{q^2}-g_{\mu\nu}\right)(\slashed{q}+M_X)\Pi^{3/2}(q^2)+\cdots,\\
\nonumber\Pi_{\mu\nu\alpha\beta}(q^2)&=&i\int d^4x e^{iq\cdot x}\langle 0|T\left[J_{\mu\nu}(x)\bar{J}_{\alpha\beta}(0)\right]|0\rangle\\
\nonumber&=&\left(g_{\mu\alpha}g_{\nu\beta}+g_{\mu\beta}g_{\nu\alpha}\right)(\slashed{q}+M_X)\Pi^{5/2}(q^2)+\cdots.
\end{eqnarray}
where $\cdots$ contains the other coupling states, $M_X$ denotes the mass of physical state $X$. In this work, we will only use the structures $1$, $g_{\mu\nu}$ and $g_{\mu\alpha}g_{\nu\beta}+g_{\mu\beta}g_{\nu\alpha}$ for the correlation functions $\Pi(p^2)$, $\Pi_{\mu\nu}(p^2)$ and $\Pi_{\mu\nu\alpha\beta}(p^2)$ respectively to study the $J^P=1/2^-,3/2^-$ and $5/2^-$ double charm pentaquark states. We assume that the current couples to the physical state $X$ through
\begin{equation}\label{parity+}
\begin{split}
 \langle 0|J|X_{1/2}\rangle&=f_Xu(p),\\
 \langle 0|J_\mu|X_{3/2}\rangle&=f_Xu_\mu(p),\\
 \langle 0|J_{\mu\nu}|X_{5/2}\rangle&=f_Xu_{\mu\nu}(p),\\
\end{split}
\end{equation}
where $f_X$ denotes the coupling constant, $u(p)$ denotes the Dirac spinor and $u_\mu(p),u_{\mu\nu}(p)$ denotes the Rarita-Schwinger vector and tensor, respectively.

\par For the convenience of discussing the parity of hadron currents, we assumed that the hadron state $X$ has the same parity as its current $J$, and used the non-$\gamma_5$ coupling relation in \eqref{parity+}. Meanwhile, the $\gamma_5$ coupling relation also exists
\begin{equation}\label{parity-}
\begin{split}
 \langle 0|J|X'_{1/2}\rangle&=f_X\gamma_5u(p),\\
 \langle 0|J_\mu|X'_{3/2}\rangle&=f_X\gamma_5u_\mu(p),\\
 \langle 0|J_{\mu\nu}|X'_{5/2}\rangle&=f_X\gamma_5u_{\mu\nu}(p),\\
\end{split}
\end{equation}
where $X'$ has the opposite parity of $X$. Eqs.\eqref{parity+},\eqref{parity-} indicate the fact that two states with opposite parity could couple to the same current . These relations also suggest that the current $j\equiv\gamma_5 J$ with opposite parity can couple to the state $X$. In the following discussion we will denote the currents with positive parity as $J$ and the currents with negative parity as $j$. The parity issue will be further discussed at the end of this section.
\par At the hadron level, two-point correlation function can be written as
\begin{equation}
\Pi(q^2)=\frac{1}{\pi}\int^{\infty}_{s_<}\frac{\mathrm{Im}\Pi(s)}{s-q^2-i\epsilon}ds,
\end{equation}
where we have used the form of the dispersion relation, and $s_<$ denotes the physical threshold. The imaginary part of the correlation function is defined as the spectral function, which is usually evaluated at the hadron level by inserting intermediate hadron states $\sum_n|n\rangle\langle n|$
\begin{equation}\label{spectral}
\begin{split}
\rho(s)\equiv\frac{1}{\pi}\mathrm{Im}\Pi(s)&=\sum_n\delta(s-M^2_n)\langle 0|J|n\rangle\langle n|\bar{J}|0\rangle\\
&=f^{-2}_X(\slashed{p}+m_X^-)\delta(s-m^{-2}_X)+f^{+2}_X(\slashed{p}-m_X^+)\delta(s-m^{+2}_X)+\mathrm{continuum}.
\end{split}
\end{equation}
The spectral density $\rho(s)$ can also be evaluated at the quark-gluon level via the operator product expansion(OPE). After performing the Borel transform at both the hadron and quark-gluon levels, the two-point correlation function can be expressed as
\begin{equation}\label{Borel}
\Pi(M^2_B)\equiv\mathcal{B}_{M^2_B}\Pi(p^2)=\int^{\infty}_{s_<}e^{-s/M^2_B}\rho(s)ds.
\end{equation}
Finally, we assume that the contribution from the continuum states can be approximated well by the OPE spectral density above a threshold value $s_0$ (duality), and arrive at the sum rule relation which can be used to perform numerical analysis:
\begin{equation}\label{QCDSRd}
M^2_X(s_0,M_B)=\frac{\int^{s_0}_{s_<}e^{-s/M^2_B}\rho(s)sds}{\int^{s_0}_{s_<}e^{-s/M^2_B}\rho(s)ds}.
\end{equation}
\par To further discuss the parity of the hadron states, we assume that the correlation function of $J$ is given by
\begin{equation*}
\Pi_+(p^2)=\slashed{p}\Pi_1(p^2)+\Pi_2(p^2),
\end{equation*}
each scalar functions $\Pi_{1,2}(p^2)$ in the equation above can construct a sum rule with \eqref{QCDSRd} separately, meanwhile, the correlation function of $j$ can be written as
\begin{equation*}
\Pi_-(p^2)=\slashed{p}\Pi_1(p^2)-\Pi_2(p^2).
\end{equation*}
The difference between the correlation function of $J$ and $j$ appears only in the sign in front of $\Pi_2(p^2)$ due to the $\gamma_5$ coupling. Thus, the same functions $\Pi_1$ and $\Pi_2$ appear in $\Pi_+$ and $\Pi_-$ bring us no independent sum rule from $j$. Here we use the method of parity projected sum rule to obtain two independent sum rules with different parity\cite{Jido:1996ia}\cite{Ohtani:2012ps}.
\par In the zero-width resonance approximation, the imaginary part of the correlation function in the rest frame $\overrightarrow{p}=0$ is considered as
\begin{equation}\label{p0}
\begin{split}
\frac{\mathrm{Im}\Pi(p_0)}{\pi}&=\sum_n\left[(\lambda^+_n)^2\frac{\gamma_0+1}{2}\delta(p_0-m^+_n)+(\lambda^-_n)^2\frac{\gamma_0-1}{2}\delta(p_0-m^-_n)\right]\\
&\equiv \gamma_0p_0\rho_{A}(p_0)+\rho_B(p_0),
\end{split}
\end{equation}
where $\lambda^{\pm}$ are coupling constants and $m^{\pm}$ denote the mass of positive or negative parity state. $\rho_A(p_0),\rho_B(p_0)$ are defined by
\begin{eqnarray*}
p_0\rho_A(p_0)&\equiv&\frac{1}{2}\sum_n\left[(\lambda^+_n)^2\delta(p_0-m^+_n)+(\lambda^-_n)^2\delta(p_0-m^-_n)\right],\\
\rho_B(p_0)&\equiv&\frac{1}{2}\sum_n\left[(\lambda^+_n)^2\delta(p_0-m^+_n)-(\lambda^-_n)^2\delta(p_0-m^-_n)\right].
\end{eqnarray*}
The combination $p_0\rho_A(p_0)+\rho_B(p_0)$ and $p_0\rho_A(p_0)-\rho_B(p_0)$ contain contributions only from the positive- or negative-parity states obviously, thus we can establish the corresponding parity projected sum rules:
\begin{eqnarray}\label{EqL}
\nonumber\mathcal{L}_k(s_0^{+},M_B^2,+)&\equiv&\frac{1}{2}\int^{s_0^{+}}_{s_<}e^{-s/M^2_B}\left[\sqrt{s}\rho_A^{\mathrm{OPE}}(s)+\rho_B^{\mathrm{OPE}}(s)\right]s^kds=\lambda_+^2m_+^{2k+1}\mathrm{exp}\left[-\frac{m_+^2}{M^2_B}\right],\\
\mathcal{L}_k(s_0^{-},M_B^2,-)&\equiv&\frac{1}{2}\int^{s_0^{-}}_{s_<}e^{-s/M^2_B}\left[\sqrt{s}\rho_A^{\mathrm{OPE}}(s)-\rho_B^{\mathrm{OPE}}(s)\right]s^kds=\lambda_-^2m_-^{2k+1}\mathrm{exp}\left[-\frac{m_-^2}{M^2_B}\right].
\end{eqnarray}
where $s_0^{\pm}$ denote the threshold of positive or negative parity state. We can extract the mass for positive and negative parity states by
\begin{equation}\label{EqMass}
m_\pm(s_0^{\pm},M_B)=\sqrt{\frac{\mathcal{L}_1(s_0^{\pm},M_B^2,\pm)}{\mathcal{L}_0(s_0^{\pm},M_B^2,\pm)}}.
\end{equation}
We shall discuss the detail to obtain suitable parameter working regions in QCD sum rule analysis in next section.
\par Using the operator product expansion (OPE) method, the two-point function can also be evaluated at the quark-gluonic level as a function of various QCD parameters. To evaluate the Wilson coefficients, we adopt the quark propagator in momentum space and the propagator 
\begin{eqnarray}
 i S_{Q}^{a b}(p)&=&\frac{i \delta^{a b}}{\slashed{p}-m_{Q}}
 +\frac{i}{4} g_{s} \frac{\lambda_{a b}^{n}}{2} G_{\mu \nu}^{n} \frac{\sigma^{\mu \nu}\left(\slashed{p}+m_{Q}\right)+\left(\slashed{p}+m_{Q}\right) \sigma^{\mu \nu}}{(p^2-m_Q^2)^2}
 +\frac{i \delta^{a b}}{12}\left\langle g_{s}^{2} G G\right\rangle m_{Q} \frac{p^{2}+m_{Q} \slashed{p}}{(p^{2}-m_{Q}^{2})^{4}}, \\
\nonumber i S_{q}^{ab}(x)&=&\frac{i\delta^{ab}}{2\pi^2x^4}\slashed{x}-\frac{\delta^{ab}}{12}\langle\bar{q}q\rangle+\frac{i}{32\pi^2}\frac{\lambda^n_{ab}}{2}g_sG^n_{\mu\nu}\frac{1}{x^2}(\sigma^{\mu\nu}\slashed{x}+\slashed{x}\sigma^{\mu\nu})\\
 & &+\frac{\delta^{ab}x^2}{192}\langle\bar{q}g_s\sigma\cdot Gq\rangle-\frac{m_q\delta^{ab}}{4\pi^2x^2}+\frac{i\delta^{ab}m_q\langle\bar{q}q\rangle}{48}\slashed{x}-\frac{im_q\langle\bar{q}g_s\sigma\cdot Gq\rangle\delta^{ab}x^2\slashed{x}}{1152},
\end{eqnarray}
where $Q$ represents the heavy quark $c$ or $b$, $q$ represents the light quark $u,d,s$, the superscripts $a, b$ denote the color indices. In this work, we will evaluate Wilson coefficients of the correlation function up to dimension nine condensates at the leading order in $\alpha_s$.

\section{Numerical Analysis}\label{Numerical}
\par In this section we perform the QCD sum rule analysis for double heavy molecular pentaquark systems using the interpolating currents in Eqs.~\eqref{Eq:eta},\eqref{Eq:xi}-\eqref{Eq:J2}. We use the standard values of various QCD condensates as $\langle \bar{q}q\rangle(1\mathrm{GeV})=-(0.24\pm0.03)^3\;\mathrm{GeV}^3$, $\langle \bar{q}g_s\sigma\cdot Gq\rangle(1\mathrm{GeV})=-M_0^2\langle \bar{q}q\rangle$, $M_0^2=(0.8\pm0.2)\;\mathrm{GeV}^2$, $\langle \bar{s}s\rangle/\langle \bar{q}q\rangle=0.8\pm0.1$, $\langle g_s^2GG\rangle(1\mathrm{GeV})=(0.48\pm0.14)\;\mathrm{GeV}^4$ at the energy scale $\mu=1$GeV~\cite{Narison:1989aq,Jamin:2001zr,Jamin:1998ra,Ioffe:1981kw,Chung:1984gr,Dosch:1988vv,Khodjamirian:2011ub,Francis:2018jyb} and $m_s(2\;\mathrm{GeV})=95^{+9}_{-3}\;\mathrm{MeV}$, $m_c(m_c)=1.27^{+0.03}_{-0.04}\;\mathrm{GeV}$, $m_b(m_b)=4.18_{-0.03}^{+0.04}\;\mathrm{GeV}$ from the Particle Data Group\cite{ParticleDataGroup:2022pth}. We also take into account the energy-scale dependence of the above parameters from the renormalization group equation
\allowdisplaybreaks{
\begin{eqnarray}\label{inputparameter}
\nonumber&&m_s(\mu)=m_s(2\mathrm{GeV})\left[\frac{\alpha_s(\mu)}{\alpha_s(2\mathrm{GeV})}\right]^{\frac{12}{33-2n_f}},\\
\nonumber&&m_c(\mu)=m_c(m_c)\left[\frac{\alpha_s(\mu)}{\alpha_s(m_c)}\right]^{\frac{12}{33-2n_f}},\\
\nonumber&&m_b(m_b)=m_b(m_b)\left[\frac{\alpha_s(\mu)}{\alpha_s(m_b)}\right]^{\frac{12}{33-2n_f}},\\
\nonumber&&\langle \bar{q}q\rangle(\mu)=\langle \bar{q}q\rangle(1\mathrm{GeV})\left[\frac{\alpha_s(1\mathrm{GeV})}{\alpha_s(\mu)}\right]^{\frac{12}{33-2n_f}},\\
  &&\langle \bar{s}s\rangle(\mu)=\langle \bar{s}s\rangle(1\mathrm{GeV})\left[\frac{\alpha_s(1\mathrm{GeV})}{\alpha_s(\mu)}\right]^{\frac{12}{33-2n_f}},\\
\nonumber&&\langle \bar{q}g_s\sigma\cdot Gq\rangle(\mu)=\langle \bar{q}g_s\sigma\cdot Gq\rangle(1\mathrm{GeV})\left[\frac{\alpha_s(1\mathrm{GeV})}{\alpha_s(\mu)}\right]^{\frac{2}{33-2n_f}},\\
\nonumber&&\langle \bar{s}g_s\sigma\cdot Gs\rangle(\mu)=\langle \bar{s}g_s\sigma\cdot Gs\rangle(1\mathrm{GeV})\left[\frac{\alpha_s(1\mathrm{GeV})}{\alpha_s(\mu)}\right]^{\frac{2}{33-2n_f}},\\  
\nonumber&&\alpha_s(\mu)=\frac{1}{b_0t}\left[1-\frac{b_1}{b_0}\frac{\mathrm{log}t}{t}+\frac{b_1^2(\mathrm{log}^2t-\mathrm{log}t-1)+b_0b_2}{b_0^4t^2}\right],
\end{eqnarray}
}
where $t=\mathrm{log}\frac{\mu^2}{\Lambda^2}$, $b_0=\frac{33-2n_f}{12\pi}$, $b_1=\frac{153-19n_f}{24\pi^2}$, $b_2=\frac{2857-\frac{5033}{9}n_f+\frac{325}{27}n_f^2}{128\pi^3}$, $\Lambda=$210 MeV, 292 MeV and 332 MeV for the flavors $n_f=$5, 4 and 3, respectively. In this work, we evolve all the input parameters to the energy scale $\mu=2m_c$ for our sum rule analysis.

\par To establish a stable mass sum rule, one should find the appropriate parameter working regions at first, i.e, for the continuum threshold $s_0$ and the Borel mass $M_B^2$. The threshold $s_0$ can be determined via the minimized variation of the hadronic mass $m_X$ with respect to the Borel mass $M_B^2$. The lower bound on the Borel mass $M_B^2$ can be fixed by requiring a reasonable OPE convergence, while its upper bound is determined through a sufficient pole contribution. The pole contribution (PC) is defined as
\begin{equation}\label{EqPC}
\mathrm{PC}(s_0^{\pm},M_B^2,\pm)=\frac{\mathcal{L}_0(s_0^{\pm},M_B^2,\pm)}{\mathcal{L}_0(\infty,M_B^2,\pm)},
\end{equation}
where $\mathcal{L}_0$ has been defined in Eq.~\eqref{EqL}.
\begin{figure}[htbp]
\centering
\includegraphics[width=10cm]{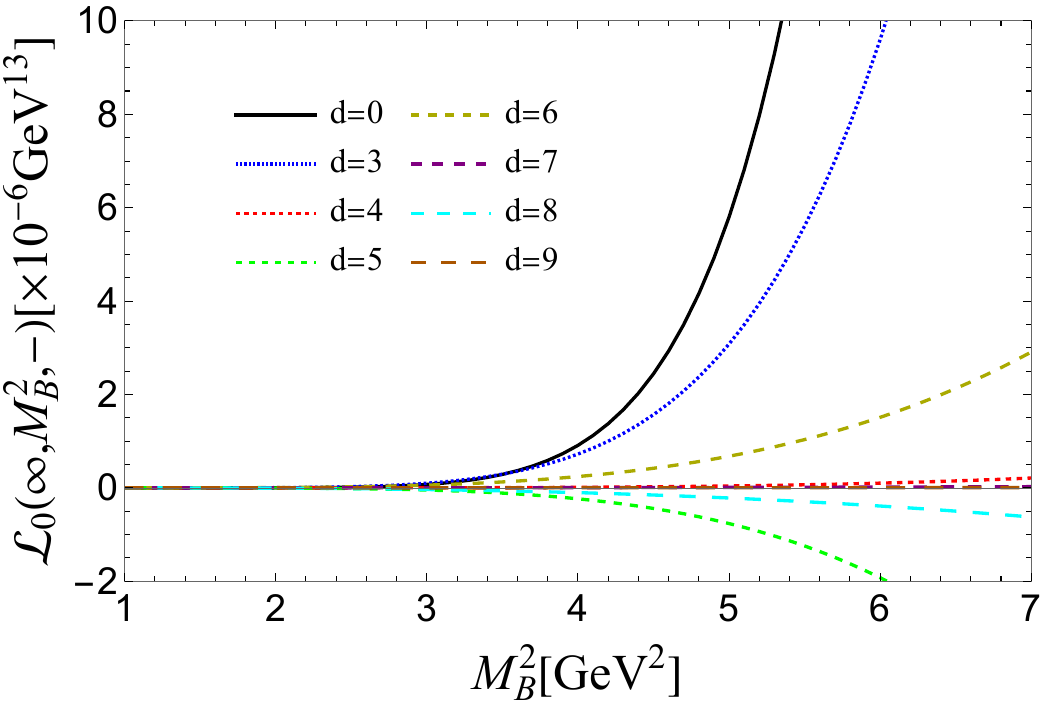}\\
\caption{Contributions of various OPE terms with dimension $d=0$ to $9$ in Eq.\eqref{EqL} for the current $J_{1,2}(x)$ with negative parity in $c$ sector, as a function of $M_B^2$ when $s_0\to\infty$.}
\label{Fig:psi4mb-rho}
\end{figure}

\begin{figure}[htbp]
\centering
\includegraphics[width=8cm]{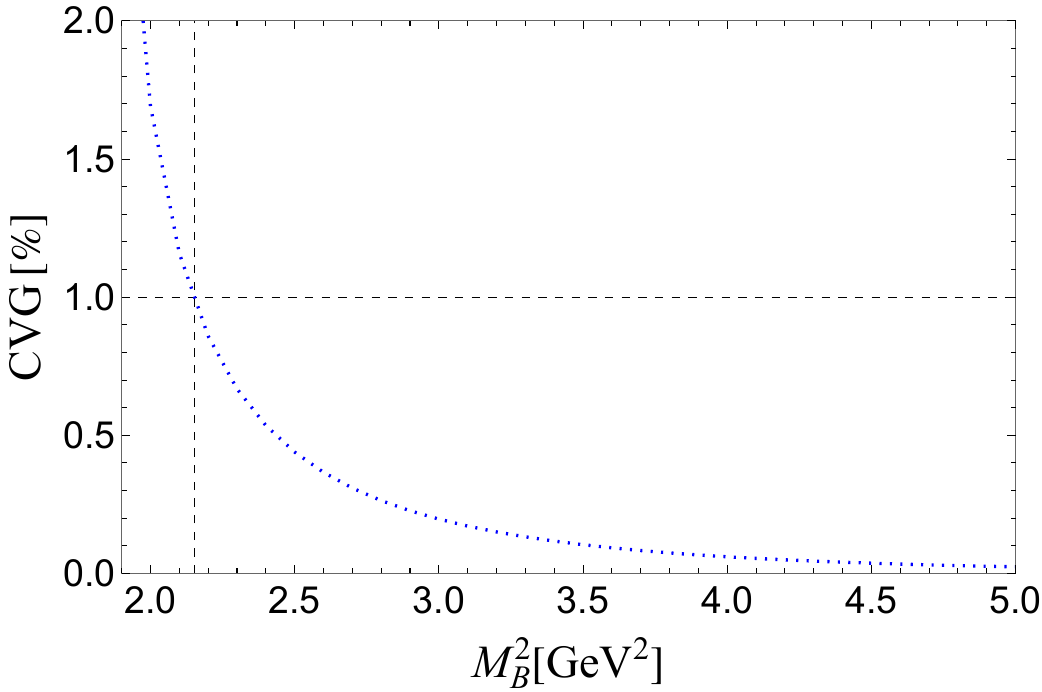}\quad
\includegraphics[width=8cm]{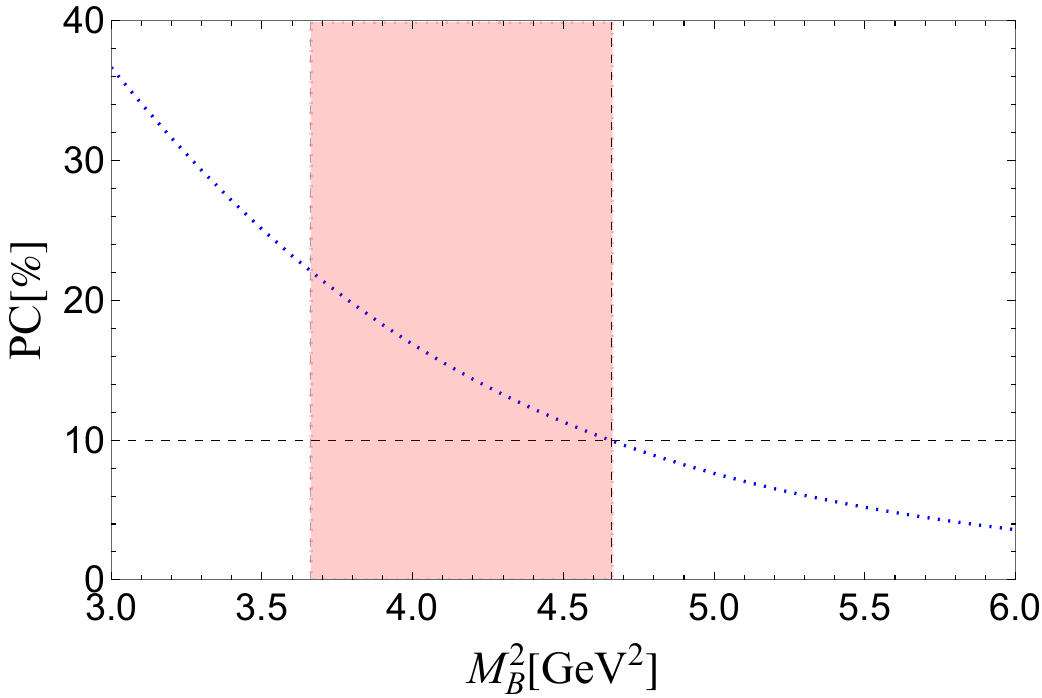}\\
\caption{Convergence(left) and Pole Contribution(right) for the interpolating current $J_{1,2}(x)$ with $J^{P}=1/2^{-}$ in $c$ sector. The band shows the working Borel window $3.47\;\mathrm{GeV}^2\leq M_B^2\leq4.47\;\mathrm{GeV}^2$.}
\label{Fig:psi4mb-CVG&PC}
\end{figure}
As an example, we use the current $J_{1,2}(x)$ with $J^{P}=1/2^{-}$ in $c$ sector to show the details of the numerical analysis. As we mainly focus on the negative parity states in this work, we will omit the parity superscript without ambiguity in the following discussion. For this current, the dominant non-perturbative contribution to the correlation function comes from the quark condensate, which is proportional to the charm quark mass $m_c$. In Fig.~\ref{Fig:psi4mb-rho}, we show the contributions of the perturbative term and various condensate terms to the correlation function with respect to $M_B^2$ when $s_0$ tends to infinity. It is clear that the Borel mass $M_B^2$ should be large enough to ensure the convergence of the OPE series. In this work, we require that the dimension-9 term makes at least 1\% contribution, which is 
\begin{equation}
\mathrm{CVG}(M_B^2,\pm)=\frac{\mathcal{L}_0^{d=9}(\infty,M_B^2,\pm)}{\mathcal{L}_0(\infty,M_B^2,\pm)}\leq1\%,
\end{equation}
providing the lower bound of the Borel mass $M_B^2\geq2.15\;\mathrm{GeV}^2$. After studying the pole contribution defined in Eq.~\eqref{EqPC}, one finds that the PC is small for pentaquark system due to the high dimension of the interpolating current. To find an upper bound of the Borel mass,  we require that the pole contribution to be larger than $10\%$. As the right panel of Fig.~\ref{Fig:psi4mb-CVG&PC} shown, the upper bound of the Borel mass is determined as 4.47 $\mathrm{GeV}^2$. With the guarantee of the validity of OPE, we choose a Borel window with a width of 1 $\mathrm{GeV}^2$. As a result, the reasonable Borel window for the current $J_{1,2}(x)$ is obtained as $3.47\;\mathrm{GeV}^2\leq M_B^2\leq4.47\;\mathrm{GeV}^2$. 
\begin{figure}[htbp]
\centering
\includegraphics[width=8cm]{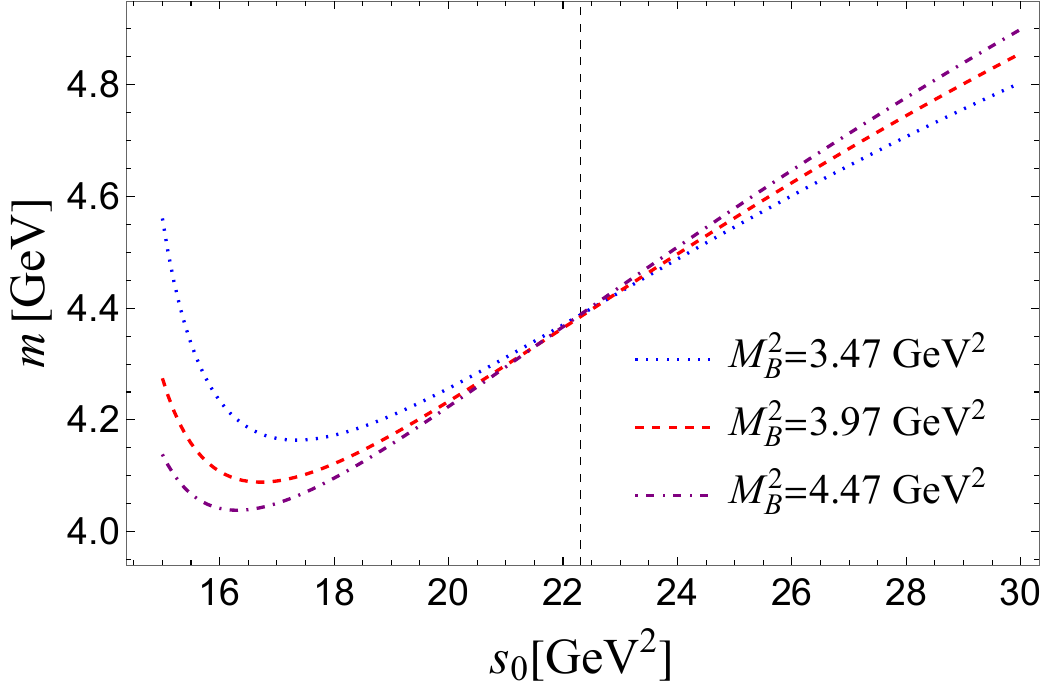}\quad
\includegraphics[width=8cm]{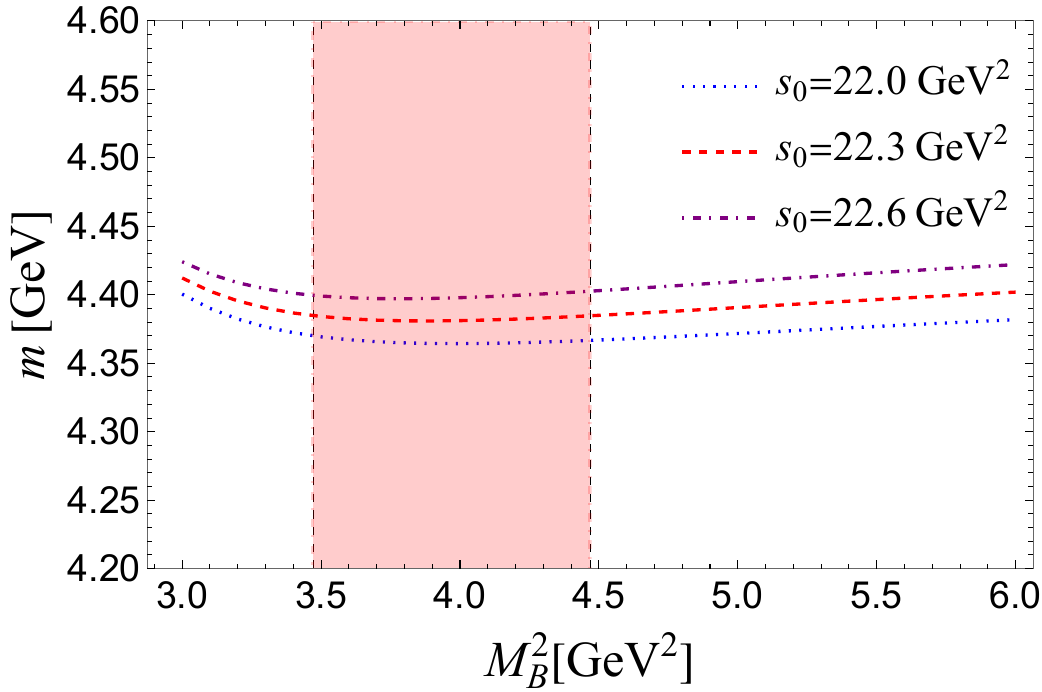}\\
\caption{Mass curves for the interpolating current $J_{1,2}(x)$ with $J^{P}=1/2^{-}$ in $c$ sector, as a function of the threshold $s_0$(left) and the Borel mass $M_B^2$(right). The band shows the working Borel window $3.47\;\mathrm{GeV}^2\leq M_B^2\leq4.47\;\mathrm{GeV}^2$. In the left panel, the mass curves with different $M_B^2$ intersect at $s_0=22.3\mathrm{GeV}^2$, indicating that the mass has a minimum dependency on the Borel mass at $s_0=22.3\mathrm{GeV}^2$. In the right panel, the mass curves with different $s_0$ around 22.3 $\mathrm{GeV}^2$ are stable in our working Borel window. The blue dotted, red dashed and purple dot-dashed curves on the left (right) figure are for Borel mass (threshold) $M_B^2=3.47~\mathrm{GeV}^2, 3.97~\mathrm{GeV}^2, 4.47~\mathrm{GeV}^2$ ($s_0=22.0~\mathrm{GeV}^2, 22.3~\mathrm{GeV}^2, 22.6~\mathrm{GeV}^2$), respectively.}
\label{Fig:psi4mb-mX}
\end{figure}

As mentioned above, the variation of the extracted hadron mass $m$ with respect to $M_B^2$ should be minimized to obtain the optimal value of the continuum threshold $s_0$. We define the following hadron mass $\bar{m}_X$ and quantity $\chi^2(s_0)$ to study the stability of mass sum rules
\begin{eqnarray}
\bar{m}(s_0,\pm)&=&\sum^N_{i=1}\frac{m(s_0,M_{B,i}^2,\pm)}{N},\\
\chi^2(s_0,\pm)&=&\sum^N_{i=1}\left[\frac{m(s_0,M_{B,i}^2,\pm)}{\bar{m}(s_0,\pm)}-1\right]^2,
\end{eqnarray}
where the $M_{B,i}^2(i=1,2,\dots,N)$ represent $N$ definite values for the Borel parameter $M_B^2$ in the Borel window. According to the above definition, the optimal choice for the continuum threshold $s_0$ in the QCD sum rule analysis can be obtained by minimizing the quantity $\chi^2(s_0)$, which is only the function of $s_0$. In this example, there is a minimum point around $s_0\approx22.3\;\mathrm{GeV}^2$ in $\chi^2$ function and we show the variation of $m$ with $s_0$ in the left panel of Fig.~\ref{Fig:psi4mb-mX}, from which we can find that the optimized value of the continuum threshold can be chosen as $s_0\approx 22.3\;\mathrm{GeV}^2$ indeed. In the right panel of Fig.~\ref{Fig:psi4mb-mX}, the mass sum rules are established to be very stable in the above parameter regions of $s_0$ and $M_B^2$. The hadron mass for this molecular pentaquark with $J^{P}=1/2^{-}$ can be obtained as 
\begin{equation}
m_{J_{1,2}}=4.38_{-0.05}^{+0.04} ~\text{GeV}\,, 
\end{equation}
where the errors come from the uncertainties of the threshold $s_0$, Borel mass $M_B^2$, quark masses and various QCD condensates. Performing the same numerical analysis to all interpolating currents in Eqs.~\eqref{Eq:eta},\eqref{Eq:xi}-\eqref{Eq:J2}, we collect their numerical results with stable sum rule analysis in Table~\ref{Tab:Result1-}-\ref{Tab:Result2-}.
\par Furthermore, we consider the dependence of the mass of the double heavy pentaquark on the mass of the heavy quark by varying the heavy quark mass to perform the sum rules analysis. In heavy quark spin symmetry, the mass of heavy hadron can be written as follows~\cite{Neubert:1993mb,Luke:1990eg,Falk:1992fm},
\begin{equation}\label{Eq:massinHQSS1}
m_{P_{QQ}}=2m_Q+\bar{\Lambda}+\frac{\Delta m^2}{4 m_Q}+O(1/m_Q^2),
\end{equation}
where $\bar{\Lambda}$ denotes the contribution independent with heavy quark mass and spin. $\Delta m^2$ denotes the contribution from heavy quark spin symmetry breaking of the order $1/m_Q$. We choose 10 testing points with masses equidistant from $m_c$ to $m_b$, and fit our results using
\begin{equation}\label{Eq:massinHQSS2}
m_{P_{QQ}}=2\,m_Q+b+\frac{c}{m_Q}.
\end{equation}
For example, we show the dependence of the pentaquark mass on the heavy quark mass from current $J_{1,2}$ in Fig.~\ref{Fig:MQrunning}.
\begin{figure}[htbp]
\centering
\includegraphics[width=10cm]{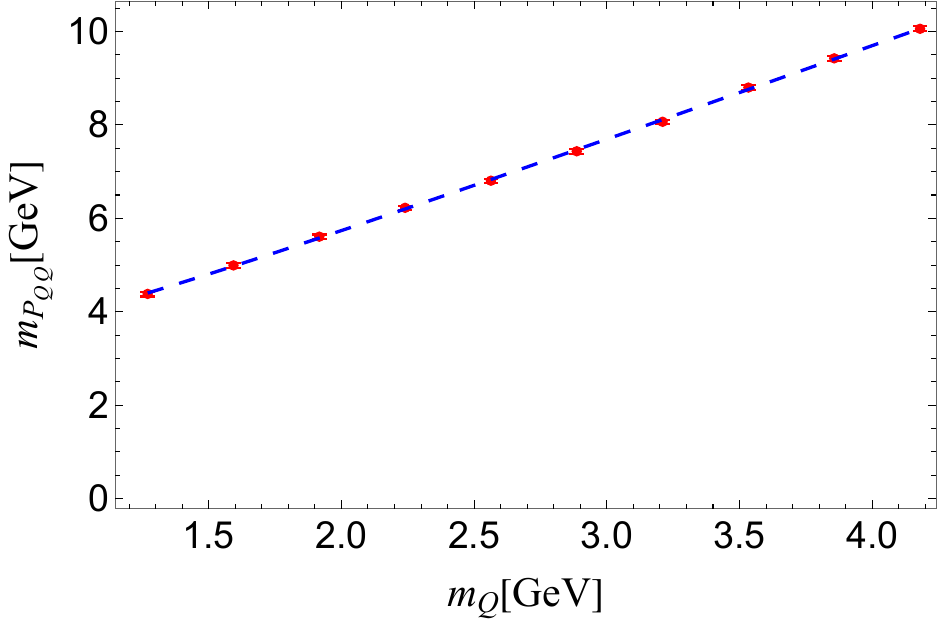}\\
\caption{ The linear dependence of double heavy pentaquarks on the heavy quark mass $m_Q$. The red dots are the QCD sum rule results with errors from the Borel windows, the threshold of higher states, the various condensates and the quark masses. The blue dashed curve is the fitted result with the formula in Eq.~\eqref{Eq:massinHQSS2}. Here the current $J_{1,2}$ is presented as an illustration. The cases for other currents are also linear.}
\label{Fig:MQrunning}
\end{figure}
We collect all the fitting parameters in Eq.~\eqref{Eq:massinHQSS2} and the mass of the bottom partner for currents we dealt with in Table~\ref{Tab:HQSS1}-\ref{Tab:HQSS2}.
\section{Discussion and Conclusion}\label{Resultanddis}
\par We have investigated the mass spectra for $[\Xi_c^{(\ast)+}D^{(\ast)+}]$, $[\Xi_{cc}^{(\ast)+}\bar{K}^{(\ast)+}]$, $[\Omega_{cc}^{(\ast)+}\pi^{+}(\rho^{+})]$ molecular pentaquark states and $[cu][cs]\bar{d}$ ,$[cc][us]\bar{d}$ compact pentaquark states in the framework of QCD sum rule. We construct the interpolating pentaquark currents and calculate their two-point correlation functions including perturbative term and various condensate terms. With appropriate Borel mass and threshold, we obtain stable sum rule for some currents and extract the corresponding mass spectra listed in Table~\ref{Tab:Result1-} and Table~\ref{Tab:Result2-}, as well as those in Fig~\ref{Fig:spectra}. In Fig.~\ref{Fig:spectra}, the two-hadron thresholds are also plotted to illustrate whether the pentaquarks are stable or not. Here the $\Xi_{cc}$ and $\Omega_{cc}$ masses are taken from the Lattice calculation in Refs.\cite{Perez-Rubio:2015zqb}.
We also list their possible strong decay mode in Table~\ref{Tab:Decay}, where we denote the negative parity baryon state $H$ with spin-$J$ as $H(J^-)$. The production for double charm pentaquark $P_{cc}$ has already been discussed in Ref.\cite{Xing:2021yid,Duan:2024uuf}. The production mechanism of strange double charm pentaquark $P_{ccs}$ is similar, i.e. via the weak decay of double heavy baryon $\Xi_{bc}$ or triply charm baryon $\Omega_{ccc}$.

\begin{figure}[htbp]
\centering
\includegraphics[width=10cm]{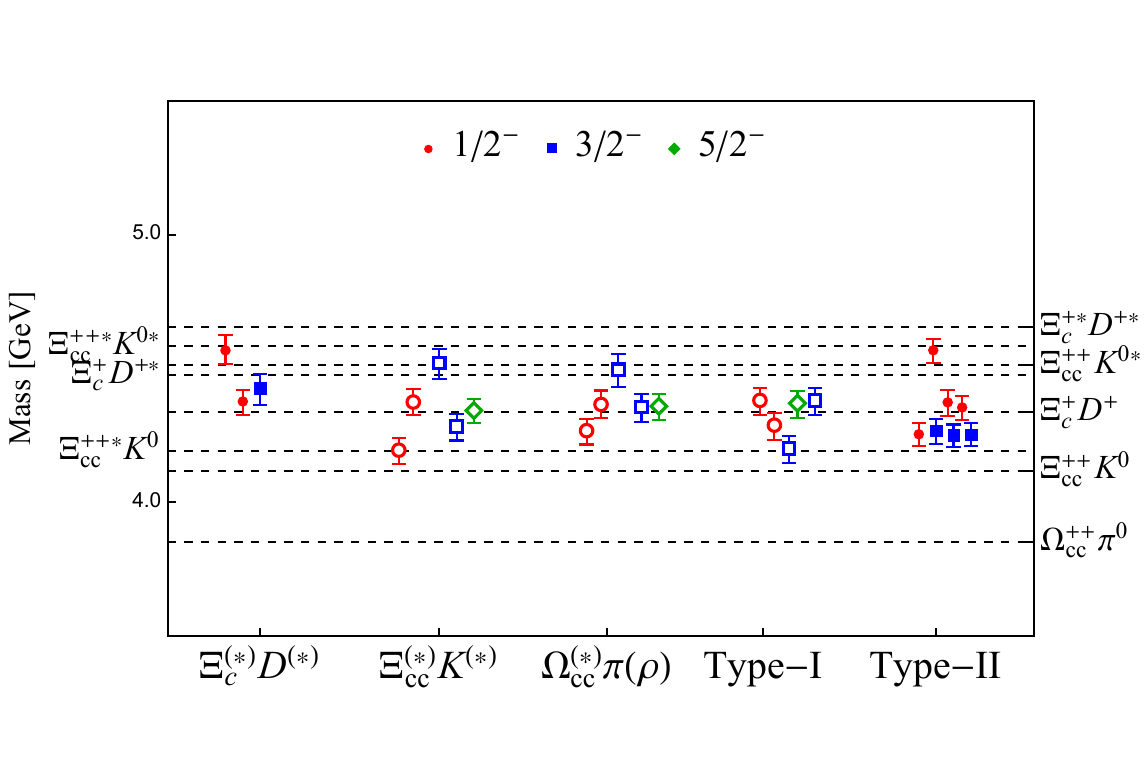}\\
\caption{The mass spectrum of strange double charm pentaquark for the interpolating currents, $\Xi_cD$, $\Xi_{cc}K$, $\Omega_{cc}\pi$, Type-I and Type-II currents in order. The red circles, blue boxes, red rhomboids are $J^P=\frac{1}{2}^-$, $\frac{3}{2}^-$ and $\frac{5}{2}^-$, respectively. The errors are from the Borel windows, the threshold of higher states, the various condensates and the quark masses. The hollow points represent the HADS partner of single charm tetraquark. 
The eight horizontal dashed lines are the double charm two-hadron thresholds. }
\label{Fig:spectra}
\end{figure}
\begin{table}[htbp]
\caption{The masses (the last second column) of various molecular structures extracted from the two-point correlation function based on QCD sum rule approach. The errors are from the Borel windows, the threshold of higher states, the various condensates and the quark masses. The first five columns present the names of the currents, the corresponding molecular structure, the quantum number $J^P$, the Borel mass square and the threshold of the continuum $s_0$. The last column indicates the corresponding two-heavy-hadron thresholds.
}\label{Tab:Result1-}\renewcommand\arraystretch{1.6} 
\setlength{\tabcolsep}{0.4 em}{ 
\centering
\begin{tabular}{c c c c c c c}
  \hline
 \hline
Current & Structure & $J^P$ & $M_B^2[\mathrm{GeV}^2]$ & $s_0[\mathrm{GeV}^2]$ & Mass[GeV] & Threshold[MeV] \\ 
 \hline
 $\eta_3$   & $\Xi^{+}_{c} D^{\ast +}$ & $\frac{1}{2}^-$ & 4.31 & 22.3 & $4.38^{+0.05}_{-0.04}$ & 4477\\
 $\eta_{4\mu}$   & $\Xi^{'+}_{c} D^{\ast +}$ & $\frac{3}{2}^-$ & 4.16 & 22.3 & $4.43^{+0.05}_{-0.04}$ & 4588\\
 $\eta_6$   & $\Xi^{\ast +}_{c} D^{\ast +}$ & $\frac{1}{2}^-$ & 3.58 & 24.5 & $4.56^{+0.06}_{-0.05}$ & 4655\\

 $\xi_1$   & $\Xi^{++}_{cc} \bar{K}^0$ & $\frac{1}{2}^-$ & 4.19 & 20.3 & $4.20^{+0.05}_{-0.05}$ & 4120\\
 $\xi_{2\mu}$   & $\Xi^{++}_{cc} \bar{K}^{\ast 0}$ & $\frac{3}{2}^-$ & 3.89 & 24.3 & $4.52^{+0.06}_{-0.05}$ & 4512\\
 $\xi_{3\mu}$   & $\Xi^{\ast ++}_{cc} \bar{K}^0$ & $\frac{3}{2}^-$ & 4.13 & 21.3 & $4.28^{+0.05}_{-0.05}$ & 4192\\
 $\xi_4$   & $\Xi^{\ast ++}_{cc} \bar{K}^{\ast 0}$ & $\frac{1}{2}^-$ & 3.56 & 22.3 & $4.37^{+0.05}_{-0.05}$ & 4584\\
 $\xi_{5\mu\nu}$   & $\Xi^{\ast ++}_{cc} \bar{K}^{\ast 0}$ & $\frac{5}{2}^-$ & 4.16 & 22.3 & $4.34^{+0.05}_{-0.04}$ & 4584\\
 $\psi_{1}$   & $\Omega^{+}_{cc} \pi^{+}$ & $\frac{1}{2}^-$ & 4.16 & 21.3 & $4.27^{+0.05}_{-0.05}$ & 3853\\
 $\psi_{2\mu}$   & $\Omega^{+}_{cc} \rho^{+}$ & $\frac{3}{2}^-$ & 3.69 & 24.3 & $4.50^{+0.06}_{-0.06}$ & 4488\\
 $\psi_{3\mu}$   & $\Omega^{\ast +}_{cc} \pi^{+}$ & $\frac{3}{2}^-$ & 4.09 & 22.3 & $4.36^{+0.05}_{-0.05}$ & 3925\\
 $\psi_{4}$   & $\Omega^{\ast +}_{cc} \rho^{+}$ & $\frac{1}{2}^-$ & 3.58 & 22.3 & $4.37^{+0.05}_{-0.05}$ & 4560\\ 
 $\psi_{5\mu\nu}$   & $\Omega^{\ast +}_{cc} \rho^{+}$ & $\frac{5}{2}^-$ & 4.38 & 22.3 & $4.36^{+0.05}_{-0.05}$ & 4560\\ 
  \hline
  \hline
\end{tabular}}
\end{table}

\begin{table}[htbp]
\caption{The masses (the last column) of various compact diquark-diquark-antiquark structures extracted from the two-point correlation function based on QCD sum rule approach. The errors are from the Borel windows, the threshold of higher states, the various condensates and the quark masses. The first five columns present the names of the currents, the corresponding diquark-diquark-antidiquark structure, the quantum number $J^P$, the Borel mass square and the threshold of the continuum $s_0$. 
The notation $[cc]_{s_1}[us]_{s_2}\bar{d}$ or $[cu]_{s_1}[cs]_{s_2}\bar{d}$ denotes the diquark components with spin $s_1$ and $s_2$, respectively. }\label{Tab:Result2-}\renewcommand\arraystretch{1.6} 
\setlength{\tabcolsep}{0.4 em}{ 
\centering
\begin{tabular}{c c c c c c}
  \hline
 \hline
Current & Structure & $J^P$ & $M_B^2[\mathrm{GeV}^2]$ & $s_0[\mathrm{GeV}^2]$ & Mass[GeV]  \\ 
 \hline
% $J_{1,1}$   & $[cc]_0[us]_0\bar{d}$ & $\frac{1}{2}^-$ & 4.16 & 21.3 & $4.28^{+0.05}_{-0.04}$ \\
 $J_{1,2}$   & $[cc]_1[us]_1\bar{d}$ & $\frac{1}{2}^-$ & 3.97 & 22.3 & $4.38^{+0.04}_{-0.05}$  \\
 $J_{1,3}$   & $[cc]_1[us]_0\bar{d}$ & $\frac{1}{2}^-$ & 4.13 & 21.3 & $4.29^{+0.05}_{-0.05}$  \\
% $J_{1,4}$   & $[cc]_0[us]_1\bar{d}$ & $\frac{1}{2}^-$ & 4.22 & 24.3 & $4.58^{+0.05}_{-0.05}$  \\
 $J_{1,5\mu}$   & $[cc]_1[us]_0\bar{d}$ & $\frac{3}{2}^-$ & 4.19 & 20.3 & $4.20^{+0.05}_{-0.05}$  \\
% $J_{1,6\mu}$   & $[cc]_0[us]_1\bar{d}$ & $\frac{3}{2}^-$ & 4.41 & 22.3 & $4.38^{+0.05}_{-0.05}$  \\
% $J_{1,7\mu}$   & $[cc]_0[us]_0\bar{d}$ & $\frac{3}{2}^-$ & 4.16 & 21.3 & $4.28^{+0.05}_{-0.04}$  \\
 $J_{1,8\mu}$   & $[cc]_1[us]_1\bar{d}$ & $\frac{3}{2}^-$ & 3.97 & 22.3 & $4.38^{+0.05}_{-0.05}$  \\ 
 $J_{1,9\mu\nu}$   & $[cc]_1[us]_1\bar{d}$ & $\frac{5}{2}^-$ & 4.16 & 22.3 & $4.37^{+0.05}_{-0.05}$  \\

 $J_{2,1}$   & $[cu]_0[cs]_0\bar{d}$ & $\frac{1}{2}^-$ & 4.47 & 21.3 & $4.25^{+0.04}_{-0.04}$  \\
 $J_{2,2}$   & $[cu]_1[cs]_1\bar{d}$ & $\frac{1}{2}^-$ & 4.88 & 24.3 & $4.57^{+0.05}_{-0.04}$  \\
 $J_{2,3}$   & $[cu]_1[cs]_0\bar{d}$ & $\frac{1}{2}^-$ & 4.47 & 22.3 & $4.37^{+0.05}_{-0.04}$  \\
 $J_{2,4}$   & $[cu]_0[cs]_1\bar{d}$ & $\frac{1}{2}^-$ & 4.25 & 22.3 & $4.35^{+0.05}_{-0.04}$  \\
 $J_{2,5\mu}$   & $[cu]_1[cs]_0\bar{d}$ & $\frac{3}{2}^-$ & 4.47 & 21.3 & $4.27^{+0.05}_{-0.04}$  \\
 $J_{2,6\mu}$   & $[cu]_0[cs]_1\bar{d}$ & $\frac{3}{2}^-$ & 4.50 & 21.3 & $4.25^{+0.04}_{-0.04}$  \\
 $J_{2,7\mu}$   & $[cu]_0[cs]_0\bar{d}$ & $\frac{3}{2}^-$ & 4.47 & 21.3 & $4.25^{+0.04}_{-0.04}$  \\
 
 \hline
  \hline
\end{tabular}}
\end{table}

\par The masses of strange double charm pentaquarks for the molecular and compact pentaquark currents are listed in Table~\ref{Tab:Result1-} and Table~\ref{Tab:Result2-}, respectively. In the two tables, only the quantum numbers $J^P=1/2^-,3/2^-,5/2^-$ are considered, as these quantum numbers can be achieved by the considered two-hadron channel in $S$-wave. The mass region for these three quantum numbers are $4.2-4.6$ GeV, $4.2-4.5$ GeV and $4.3$ GeV respectively. One should notice that the masses of the currents $\Xi_c^+D^{\ast +}$, $\Xi_c^{'+}D^{\ast +}$, $\Xi_{c}^{\ast +}D^{\ast +}$, $\Xi_{cc}^{\ast ++}\bar{K}^{\ast 0}$ and $\Omega_{cc}^{\ast +}\rho^{+}$
are below their corresponding threshold, indicating that these strange double charm pentaquarks are stable against their strong decay channels. On the contrary, those for the  
 $\Xi_{cc}^{++}\bar{K}^0$, $\Xi_{cc}^{++}\bar{K}^{\ast 0}$, $\Xi_{cc}^{\ast ++}\bar{K}^0$, $\Omega_{cc}^{+}\rho^{+}$ and $\Omega_{cc}^{\ast +}\pi^{+}$ channels are around or above their corresponding thresholds, indicating that they could be broader states and not easy to be detected in experiment. In comparison with other works in the market, we plot Fig.~\ref{Fig:otherwork}. In the figure, we also plot the thresholds of the potential strong decay channels for each quantum number. We mainly compare our results with those of Refs.~\cite{Zhou:2018bkn,Dong:2021bvy,Xing:2021yid,Zhu:2019iwm,Wang:2023eng}. 
In Ref.~\cite{Zhou:2018bkn}, the authors employ a color-magnetic interaction in Schr\"odinger equation and obtain the mass spectra by variational method. Their masses, namely the mass for $J^P=1/2^-,3/2^-,5/2^-$ are around 4.0-4.8 GeV, 4.1-4.8 GeV and 4.7 GeV respectively, are higher than ours. There are two reasons. One is that they mainly focus on the mass splitting of the pentaquark states and the accurate values need further dynamical calculations, as claimed by the authors in their work. Thus the wider range of mass in Ref.~\cite{Zhou:2018bkn} is probably due to the overlooked dynamical calculation. Another reason is that they use the variational method, which is well known that it can only gives the upper limit of a given state.  Similarly, in Ref.\cite{Park:2018oib}, the authors analyse the pentaquark $q^4\bar{Q}$ system in a constituent quark model based on the chromomagnetic interaction in both the SU(3) flavor symmetric and SU(3) flavor broken case. They find that the $\Xi_{cc}\bar{K}$ could be stable pentaquark state.
In Ref.~\cite{Zhu:2019iwm}, the authors study the double-heavy pentaquarks in non-relativistic constituent quark model by solving the multi-body Schr\"odinger equation including the color Coulomb interaction and spin-dependent interaction. They conclude that the mass about 4.4-4.5 GeV for compact pentaquark. Such results are slightly higher than our results, which may possibly due to the overlooked linear confinement potential and induce two free parameters $\beta$ and the distance between the two heavy quarks $R$. 
The authors suggest that the optimal value of $R=1 \mathrm{fm}$ for double heavy pentaquark, which may cause a higher mass spectra. Another potential reason is that they use variational method as discussed above. In Ref.~\cite{Xing:2021yid}, the authors employ the double heavy diquark-triquark model including the interaction between triquark and heavy diquark $[cc]_{\bar{3}}$ and the interaction in the light diquark $[qq^{'}]_{\bar{3}}$. They obtain the mass for $cc\bar{n}sn$ states with $J^P=1/2^-$ and $3/2^-$ as $4.1\pm0.3$ GeV and $4.6\pm0.3$ GeV, which is consistent with ours. In Ref.~\cite{Dong:2021bvy}, the authors study the heavy-heavy hadronic molecules by solving the single channel Bethe-Salpeter equation with the interactions following the heavy quark spin symmetry, including the interactions from light vector meson exchange. Their results illustrate that $D^{(*)}\Xi_c^{('*)}$ system can be bound easily, which is consistent with our results. In Ref.~\cite{Wang:2023eng}, the authors study the molecular pentaquarks by solving the Lippmann-Schwinger equations with near-threshold effective potentials. Their results suggest that $\Xi_cD^{(')\ast}$, $\Xi_c^{\ast}D^{\ast}$ and $\Xi_{cc}^{\ast}\bar{K}^{\ast}$ systems could possibly bound as molecular states, which is consistent with our results.
 
 \par It is interesting to find that we obtain almost degenerate mass for all the currents with $J^P=5/2^-$, as shown in Fig.~\ref{Fig:otherwork}. This is because that all these currents contain spin 1 $cc$ diquark, leaving potential small splitting from light quarks. In addition, one can also see that the mass is 
 far below the thresholds of corresponding strong decay channels and can be viewed as a stable and narrow state. 
As the result, we consider the existence of this state is a very solid conclusion of our work. The best observed channel is its semi-leptonic decay to double charm baryon, i.e. $\Xi_{cc}^{*++}$ and $\Omega_{cc}^{*+}$.

\begin{figure}[htbp]
\centering
\includegraphics[width=10cm]{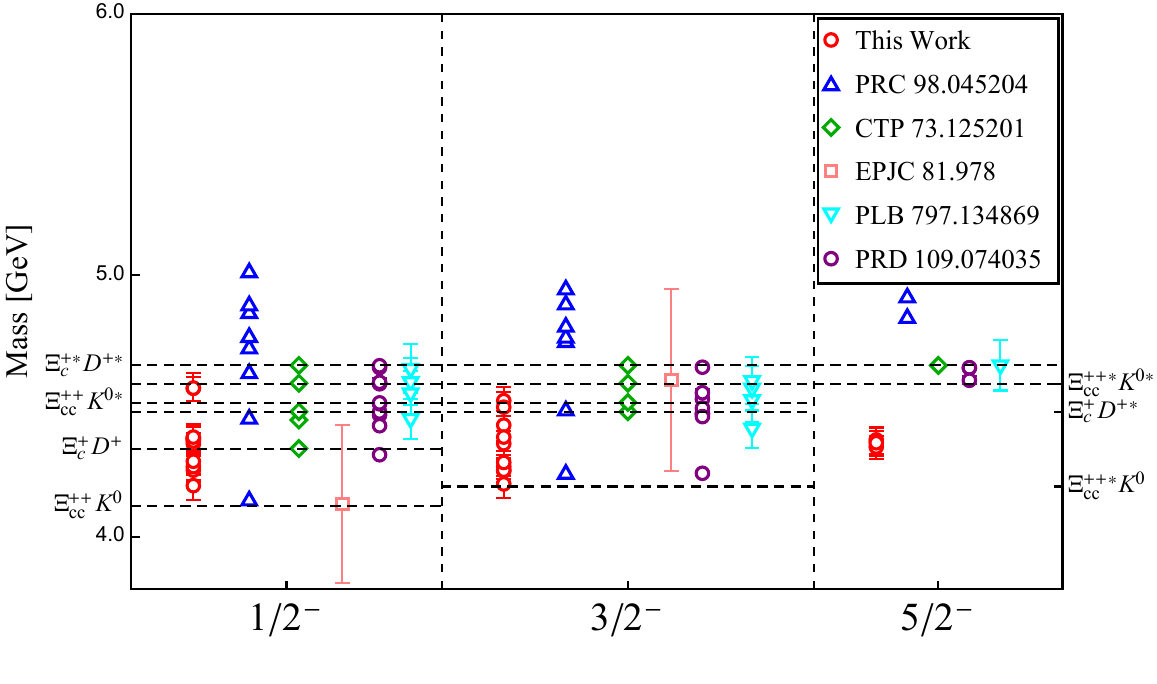}\\
\caption{Mass spectra of double charm pentaquark with $J^P=1/2^-,3/2^-,5/2^-$, in comparison with other works. The blue upside triangles, green diamonds, pink squares, cyan downside triangles, and the purple circles are the results from Ref.~\cite{Zhou:2018bkn}, Ref.~\cite{Dong:2021bvy} (cutoff $\Lambda=0.5~\mathrm{GeV}$), Ref.~\cite{Xing:2021yid}, 
Ref.~\cite{Zhu:2019iwm}, and Ref.~\cite{Wang:2023eng}, respectively. The red circles are our results. The horizontal dashed lines represents the corresponding two-hadrons thresholds for given quantum numbers.}
\label{Fig:otherwork}
\end{figure}

\par We consider the dependency of the pentaquark mass $P_{QQ}$ on the heavy quark mass $m_Q$ and use Eq.~\eqref{Eq:massinHQSS2} to fit our results. The fitting parameters and the predicted double bottom pentaquarks masses are listed in Table~\ref{Tab:HQSS1}-\ref{Tab:HQSS2}. The masses of double bottom pentaquarks are around 9.6-10.6 GeV, 10.0-10.2 GeV and 10.0 GeV for $J^P=1/2^-,3/2^-,5/2^-$ respectively. Except that the two states with $J^P=1/2^-$ are higher than the lowest two-hadron threshold $\Omega_{bb}\pi$ (10.33 GeV), all the other double bottom pentaquark states are lower than their corresponding lowest two-hadron thresholds and can be viewed as narrow states. Here the masses of double bottom baryons are taken from the result of Lattice QCD in Ref.~\cite{Mohanta:2019mxo}. From the two tables, one can also see the heavy quark spin symmetry emerging in the spectrum. As we know, spin-interaction of heavy quarks does not occur at the leading order in the $\Lambda_{\mathrm{QCD}}/m_Q$ expansion, which makes the masses of pentaquarks with the same light quark spin are degenerate.
Such symmetry can be seen in our results, namely $J_{1,2}$, $J_{1,8\mu}$ and $J_{1,9\mu\nu}$ with spin-1 $[us]$ diquark component give a degenerate mass at 4.38 GeV. Such behavior can also be seen in the molecular structure currents $\xi_1$, $\xi_{3\mu}$ in the $\Xi_{cc}^{++}\bar{K}^0$ structure and $\psi_{4}$, $\psi_{5\mu\nu}$ in the $\Omega_{cc}^{+}\rho^{+}$ structure.
Comparing Eq.~\eqref{Eq:massinHQSS1} with Eq.~\eqref{Eq:massinHQSS1}, the parameter $b$ in Eq.~\eqref{Eq:massinHQSS2} should be the parameter $\bar{\Lambda}$ in Eq.~\eqref{Eq:massinHQSS1} and is independent on heavy quark mass and spin. This feature can be reflected by the currents $(\psi_1,\psi_{3\mu})$, $(J_{1,3},J_{1,5\mu})$ and $(J_{1,2},J_{1,8\mu},J_{1,9\mu\nu})$. They have the same light quark spin structure, leaving almost the same parameter $b$.
\begin{table}[htbp]
\caption{Fitting coefficients of Eq.\eqref{Eq:massinHQSS2} for molecular pentaquarks. The first, second and third column show the currents, $J^P$ and molecular structure for each pentaquark, respectively. The fourth and the fifth columns show present the fitted coefficients $b$ and $c$. The sixth column shows the mass spectra for double bottom pentaquarks by replacing the charm quark mass by the bottom quark mass. The errors are from the Borel windows, the threshold of higher states, the various condensates and the quark masses. The last two columns list the mass and spin-parity of the HADS partners for the corresponding pentaquarks. While "$-$" denotes that there are not the HADS partner, as the two heavy quarks are not formed as a $\bar{3}$ diquark in color space.  }\label{Tab:HQSS1}\renewcommand\arraystretch{1.6} 
\setlength{\tabcolsep}{0.4 em}{ 
\centering
\begin{tabular}{c c c c c c|c c}
  \hline
 \hline
Current & $J^P$ & Structure  & $b$[GeV] & $c\mathrm{[GeV^2]}$& $m_{P_{bbs}}$[GeV] & $m_{T_{c\bar{s}}}$[GeV] & $J^P(T_{c\bar{s}})$ \\ 
 \hline
 $\eta_3$ & $\frac{1}{2}^-$ & $\Xi^{+}_{c} D^{\ast +}$  & 1.53 & 0.40& $10.04^{+0.06}_{-0.05}$ & -- & --\\
 $\eta_{4\mu}$ & $\frac{3}{2}^-$ & $\Xi^{'+}_{c} D^{\ast +}$  & 1.67 & 0.25& $10.19^{+0.06}_{-0.05}$ & --& --\\
 $\eta_6$& $\frac{1}{2}^-$  & $\Xi^{\ast +}_{c} D^{\ast +}$  & 2.26 & -0.35& $10.61^{+0.06}_{-0.05}$ & --& --\\

 $\xi_1$& $\frac{1}{2}^-$  & $\Xi^{++}_{cc} \bar{K}^0$  & 1.48 & 0.24 & $9.93^{0.06}_{0.05}$ &2.94& $0^+$\\
 $\xi_{2\mu}$ & $\frac{3}{2}^-$ & $\Xi^{++}_{cc} \bar{K}^{\ast 0}$  & 1.64 & 0.44 & $10.17^{0.06}_{0.06}$ &3.26& $1^+$\\
 $\xi_{3\mu}$ & $\frac{3}{2}^-$ & $\Xi^{\ast ++}_{cc} \bar{K}^0$  & 1.46 & 0.38 & $9.96^{+0.06}_{-0.05}$ &3.03& $1^+$\\
 $\xi_4$ & $\frac{1}{2}^-$ & $\Xi^{\ast ++}_{cc} \bar{K}^{\ast 0}$  & 1.04 & 1.07 & $9.59^{+0.07}_{-0.07}$ &3.16& $0^+$\\
 $\xi_{5\mu\nu}$& $\frac{5}{2}^-$  & $\Xi^{\ast ++}_{cc} \bar{K}^{\ast 0}$  & 1.50 & 0.38 & $9.99^{+0.06}_{-0.05}$ &3.07& $0^+,1^+,2^+$\\
 $\psi_{1}$& $\frac{1}{2}^-$ & $\Omega^{\ast +}_{cc} \pi^{+}$  & 1.48 & 0.30 & $9.96^{+0.06}_{-0.05}$ &2.99& $0^+$\\
 $\psi_{2\mu}$&  $\frac{3}{2}^-$ & $\Omega^{+}_{cc} \rho^{+}$  & 1.57 & 0.51 & $10.04^{+0.06}_{-0.06}$ &3.25& $1^+$\\
 $\psi_{3\mu}$ & $\frac{3}{2}^-$ & $\Omega^{\ast +}_{cc} \pi^{+}$  & 1.45 & 0.47 & $9.96^{+0.06}_{-0.05}$ &3.09& $1^+$\\
 $\psi_{4}$ & $\frac{1}{2}^-$ & $\Omega^{\ast +}_{cc} \rho^{+}$  & 1.16 & 0.88 & $9.71^{+0.06}_{-0.06}$ &3.12& $0^+$\\ 
 $\psi_{5\mu\nu}$ & $\frac{5}{2}^-$ & $\Omega^{\ast +}_{cc} \rho^{+}$  & 1.61 & 0.25 & $10.14^{+0.07}_{-0.06}$ &3.08& $0^+,1^+,2^+$\\ 
 \hline
 \hline
\end{tabular}}
\end{table}

\begin{table}[htbp]
\caption{The same as that of Table~\ref{Tab:HQSS2} but for compact pentaquark. The notation $[cc]_{s_1}[us]_{s_2}\bar{d}$ or $[cu]_{s_1}[cs]_{s_2}\bar{d}$ denotes the diquark components with spin $s_1$ and $s_2$, respectively.}\label{Tab:HQSS2}\renewcommand\arraystretch{1.6} 
\setlength{\tabcolsep}{0.4 em}{ 
\centering
\begin{tabular}{c c c c c c|c c}
  \hline
 \hline
Current & $J^P$ &Structure  & $b$[GeV] & $c\mathrm{[GeV^2]}$ & $m_{P_{bbs}}$[GeV] & $m_{T_{c\bar{s}}}$[GeV] & $J^P(T_{c\bar{s}})$ \\ 
 \hline
% $J_{1,1}$ & $\frac{1}{2}^-$ & $[cc]_0[us]_0\bar{d}$  & 1.50 & 0.30 & $9.97^{+0.06}_{-0.05}$ & -- & --\\
 $J_{1,2}$ & $\frac{1}{2}^-$ & $[cc]_1[us]_1\bar{d}$  & 1.61 & 0.30 & $10.06^{+0.06}_{-0.05}$ & 3.11 & $0^+,1^+$ \\
 $J_{1,3}$ & $\frac{1}{2}^-$ & $[cc]_1[us]_0\bar{d}$  & 1.52 & 0.29 & $10.00^{+0.06}_{-0.05}$ & 3.02 & $0^+$ \\
% $J_{1,4}$ & $\frac{1}{2}^-$ & $[cc]_0[us]_1\bar{d}$  & 1.84 & 0.22 & $10.36^{+0.06}_{-0.06}$ & -- & -- \\
 $J_{1,5\mu}$ & $\frac{3}{2}^-$ & $[cc]_1[us]_0\bar{d}$  & 1.53 & 0.17 & $9.97^{+0.06}_{-0.05}$ & 2.94 & $1^+$ \\
% $J_{1,6\mu}$ & $\frac{3}{2}^-$ & $[cc]_0[us]_1\bar{d}$  & 1.64 & 0.23 & $10.16^{+0.07}_{-0.06}$ & -- & -- \\
% $J_{1,7\mu}$ & $\frac{3}{2}^-$ & $[cc]_0[us]_0\bar{d}$  & 1.50 & 0.30 & $9.97^{+0.06}_{-0.05}$ & -- & -- \\
 $J_{1,8\mu}$ & $\frac{3}{2}^-$ & $[cc]_1[us]_1\bar{d}$  & 1.61 & 0.30 & $10.06^{+0.06}_{-0.05}$ & 3.11 & $1^+$ \\ 
 $J_{1,9\mu\nu}$ & $\frac{5}{2}^-$ & $[cc]_1[us]_1\bar{d}$  & 1.60 & 0.29 & $10.10^{+0.06}_{-0.06}$ & 3.09 & $2^+$ \\

 $J_{2,1}$ & $\frac{1}{2}^-$ & $[cu]_0[cs]_0\bar{d}$  & 1.79 & -0.16 & $10.21^{+0.07}_{-0.06}$ & -- & -- \\
 $J_{2,2}$ & $\frac{1}{2}^-$  & $[cu]_1[cs]_1\bar{d}$  & 1.87 & 0.14 & $10.33^{+0.07}_{-0.06}$ & -- & -- \\
 $J_{2,3}$ &  $\frac{1}{2}^-$ & $[cu]_1[cs]_0\bar{d}$  & 1.74 & 0.07 & $10.18^{+0.07}_{-0.06}$ & -- & -- \\
 $J_{2,4}$ & $\frac{1}{2}^-$  & $[cu]_0[cs]_1\bar{d}$  & 1.72 & 0.08 & $10.17^{+0.06}_{-0.06}$ & -- & -- \\
 $J_{2,5\mu}$ & $\frac{3}{2}^-$  & $[cu]_1[cs]_0\bar{d}$  & 1.65 & 0.07 & $10.08^{+0.06}_{-0.06}$ & -- & -- \\
 $J_{2,6\mu}$ & $\frac{3}{2}^-$  & $[cu]_0[cs]_1\bar{d}$  & 1.61 & 0.08 & $10.08^{+0.06}_{-0.06}$ & -- & -- \\
 $J_{2,7\mu}$ & $\frac{3}{2}^-$  & $[cu]_0[cs]_0\bar{d}$  & 1.79 & -0.16 & $10.21^{+0.07}_{-0.06}$ & -- & -- \\
 
 \hline
  \hline
\end{tabular}}
\end{table}

\par As we discussed in Sec.~\ref{Sec:Intro}, the double charm pentaquark $ccus\bar{d}$ should be the HADS partner of the singly charm tetraquark $us\bar{d}\bar{c}$, we can derive the corresponding mass spectra of singly charm tetraquark $T_{c\bar{s}}$ through Eq.~\eqref{Eq:massinHQSS1} by replacing $2m_Q$ to $m_Q$, and we list these corresponding spectra and their spin-parity in Table~\ref{Tab:HQSS1}-\ref{Tab:HQSS2}. The mass and spin-parity of HADS partner for current $\xi_1$ is consistent with the recently discovered $T_{c\bar{s}}(2900)$, which indicates that $T_{c\bar{s}}(2900)$ could be molecular tetraquark with spin-0 $[s\bar{d}]$ meson component. Furthermore, current ($J_{1,2}$, $J_{1,8\mu}$, $J_{1,9\mu\nu}$) could be the HADS partner triplet for tetraquark $[us]_1[\bar{c}\bar{d}]_1$ with mass about 3.1 GeV, current ($J_{1,3}$, $J_{1,5\mu}$) or ($\xi_1$, $\xi_{3\mu}$) could be the HADS partner doublet for tetraquark $T_{c\bar{s}}(2900)$. With the spectra and decay modes in this work, we hope that these double charm and double bottom pentaquarks could be discovered by the LHCb, BelleII, CMS and RHIC collaborations and so on in the near future. 

\begin{table}[htbp]
\caption{Possible decay mode for double charm pentaquark}\label{Tab:Decay}\renewcommand\arraystretch{1.6} 
\setlength{\tabcolsep}{0.4 em}{ 
\centering
\begin{tabular}{c c c}
  \hline
 \hline
$J^P$ & S-wave & P-wave  \\ 
 \hline
$1/2^-$ & $\Xi^{(')}_cD^{(*)}$, $\Xi_{cc}\bar{K}^{(\ast)}$, $\Omega_{cc}\pi/\rho$ & $\Xi_{cc}\bar{K}_{0}^{\ast}$, $\Xi_{cc}(1/2^-)\bar{K}$, $\Xi_{cc}^*\bar{K}^{\ast}_0$, \\
        & & $\Xi_{cc}^*(3/2^-)\bar{K}$, $\Omega_{cc}(1/2^-,3/2^-)\pi$\\
\hline
$3/2^-$ & $\Xi_cD^*$, $\Xi_c^*D$, $\Xi_{cc}\bar{K}^{\ast}$, $\Xi_{cc}^*\bar{K}$, & 
                                $\Xi_{cc}\bar{K}^{\ast}_0$, $\Xi_{cc}(1/2^-,3/2)\bar{K}$, \\
        & $\Omega_{cc}\rho$, $\Omega_{cc}^*\pi$ & $\Omega_{cc}(1/2^-,3/2^-)\pi$\\
\hline
$5/2^-$ & - & - \\
 \hline
 \hline
\end{tabular}}
\end{table}

\section{Summary}
Motivated by the observation of the $T_{c\bar{s}}(2900)$, we study the mass spectrum of its HADS counter parts, i.e. strange double charm pentaquarks. By constructing currents in the molecular and compact pentaquark pictures, we extract the corresponding mass spectra for quantum numbers $J^P=1/2^-, 3/2^-, 5/2^-$. The masses for the former two quantum numbers are within the energy region $4.2-4.6~\mathrm{GeV}$ and $4.2-4.5~\mathrm{GeV}$, respectively. The masses of the three currents (two molecular currents and one compact pentquark current) of the quantum number $J^P=5/2^-$ are degenerate and locate at $4.3~\mathrm{GeV}$. It is far below the threshold of its two-hadron strong decay channel and can be viewed as a narrow state, making it easily to be measured in experiment. The best observed channel is the semi-leptonic decay to double charm baryon. The corresponding strange double bottom pentaquarks are locate at  $9.6-10.6~\mathrm{GeV}$, $10.0-10.2~\mathrm{GeV}$ and $10.0~\mathrm{GeV}$ for the above mentioned three quantum numbers. This kind of study is useful for the further measurements of strange double charm and double bottom pentaquarks.

\section*{ACKNOWLEDGMENTS}
%We are grateful to XXX, XXX for the helpful discussion. 
This work is partly supported by the National Natural Science Foundation of China with Grant Nos.~12375073, 12035007, and 12175318, Guangdong Provincial funding with Grant Nos.~2019QN01X172, Guangdong Major Project of Basic and Applied Basic Research No.~2020B0301030008, the Natural Science Foundation of Guangdong Province of China under Grant No. 2022A1515011922, 
 the NSFC and the Deutsche Forschungsgemeinschaft (DFG, German
Research Foundation) through the funds provided to the Sino-German Collaborative
Research Center TRR110 ``Symmetries and the Emergence of Structure in QCD"
(NSFC Grant No. 12070131001, DFG Project-ID 196253076-TRR 110).

\end{document}